\documentclass[a4paper, 11pt]{article}
\pdfoutput=1

\usepackage[latin1]{inputenc}
\usepackage[T1]{fontenc}
\usepackage[sc]{mathpazo} 
\usepackage{courier}
\usepackage{float}
\usepackage[intlimits,fleqn]{amsmath}
\usepackage{amsbsy}
\usepackage{amssymb,dcolumn,exscale}
\usepackage{booktabs}
\usepackage{varioref}
\usepackage[small]{caption}
\usepackage[width=16.6cm, left=2.2cm,top=2.5cm,height=23.7cm]{geometry}
\usepackage{cite}
\usepackage{graphicx,color}
\usepackage{epsf}
\usepackage{pstricks,psfrag,booktabs,multirow,longtable,subfig}
\graphicspath{{./pic/}}

\newcommand{\as}{\alpha_s}

\newcommand{\half}{\ensuremath{\tfrac{1}{2}}}
\newcommand{\mq}{m_{\widetilde{q}}}
\newcommand{\mg}{m_{\widetilde{g}}}

\newcommand{\shat}{\hat{s}}
\newcommand{\qqbar}{\ensuremath{q \bar{q}}}
\newcommand{\gluinopair}{\ensuremath{\widetilde{g}\,\widetilde{g}}}

\newcommand{\GeV}{\ensuremath{\,\mathrm{GeV}}}
\newcommand{\TeV}{\ensuremath{\,\mathrm{TeV}}}

\newcommand{\pb}{\ensuremath{\,\mathrm{pb}}}

\newcommand{\Cqq}{C_{1}^{(\qqbar)}}
\newcommand{\Cgg}{C_{1}^{(gg)}}

\newcommand{\pptogluinopair}{\ensuremath{pp\to \gluinopair}}

\newcommand{\lat}{\widetilde\lambda}
\newcommand{\lnNt}{\ln(\widetilde N)}

\newcommand{\non}{\nonumber}
\newcommand{\ord}{{\mathcal O}}
\def\shat{{\hat s}}
\def\muf{{\mu^{}_f}}
\def\mufs{{\mu^{2}_f}}
\def\mur{{\mu^{}_r}}
\def\murs{{\mu^{2}_r}}

\DeclareMathOperator*{\Li}{Li_2}

\newcolumntype{d}[1]{D{.}{.}{#1}}

\allowdisplaybreaks[3]

\begin{document}
\begin{titlepage}
\noindent
DESY 12-142 \hfill November 2012\\
LPN 12-091 \\
SFB/CPP-12-62 \\
\vspace{2.3cm}

\noindent 
\begin{center}
\Large{\bf
QCD threshold corrections for gluino pair production \\[1ex] at hadron colliders
}\\
\vspace{1.5cm}
\large
Ulrich Langenfeld$^a$, Sven-Olaf Moch$^b$, and Torsten Pfoh$^b$\\[10mm]
\normalsize
{\it 
$^a$Julius-Maximilians-Universit\"at W\"urzburg\\[1mm]
Am Hubland, D--97074 W\"urzburg, Germany \\[2mm]
\texttt{\footnotesize{ulangenfeld@physik.uni-wuerzburg.de}}
\\[5mm]
$^b$Deutsches Elektronensynchrotron DESY\\[1mm]
Platanenallee 6, D--15738 Zeuthen, Germany \\[2mm]
\texttt{\footnotesize{sven-olaf.moch@desy.de, torsten.pfoh@desy.de}}
}
\vspace{2.3cm}

\large
{\bf Abstract}
\vspace{-0.2cm}
\end{center}
We present the complete threshold enhanced predictions in QCD 
for the total cross section of gluino pair production at hadron colliders 
at next-to-next-to-leading order.
Thanks to the computation of the required one-loop hard matching coefficients 
our results are accurate to the next-to-next-to-leading logarithm.
In a brief phenomenological study we provide predictions 
for the total hadronic cross sections at the LHC 
and we discuss the uncertainties arising from scale variations 
and the parton distribution functions.
\vfill
\end{titlepage}

\newpage
\section{Introduction}
The Minimal Supersymmetric Standard Model (MSSM) is an attractive 
extension~\cite{Haber:1984rc,Nilles:1983ge} of the very successful Standard Model (SM)
of particle physics.
One property of the MSSM is its rich spectrum of heavy particles which might
be discovered at the LHC if they are lighter than $\approx 2\TeV$.
Hadron colliders are especially appropriate to study color-charged particles.
In the MSSM, the superpartners of the gluon and the quarks are the 
gluino, which is a Majorana fermion, and the scalar quarks (squarks), respectively.
The associated superpartners of the SM particles have the same weak isospin, 
hypercharge and color charge.

Searches for supersymmetry have been performed at the Tevatron and the LHC with
center-of-mass (cms) energies of $7$ and $8\TeV$.
Due to its larger energy the best bounds on the masses of these hypothetical particles come now from the LHC. 
The production cross section for gluino pairs at the LHC is sizable. 
It is driven by the large gluon luminosity and it is 
further enhanced due to the large color charge of gluons and gluinos.
However, no superpartners of the SM particles have been discovered so far. 

A special feature of particle spectra in the  constrained MSSM (CMSSM) are the large mass 
differences between squarks, gauginos, and sleptons, allowing cascading decays of the SUSY 
particles. 
At the Atlas experiment, searches for squarks and gluinos in a CMSSM framework 
are performed by looking for final states with a large number of 
jets and missing transverse momentum~\cite{Aad:2012hm}, 
additional same sign leptons~\cite{ATLAS:2012ai} or $b$-jets~\cite{ATLAS:2012ah}.
Gluino masses smaller than $840\GeV$~\cite{Aad:2012hm}, $550-700\GeV$~\cite{ATLAS:2012ai},
and $600-900\GeV$~\cite{ATLAS:2012ah} are excluded and 
similar results~\cite{Chatrchyan:2012jx,:2012mfa,Chatrchyan:2012sa,:2012th,Chatrchyan:2012te}
are reported by the CMS experiment. 
The LHC bounds discussed above do not apply if the particle spectrum is compressed. 
In such scenarios, if the gluino is mass degenerate with the lightest supersymmetric particle 
and the squarks are decoupled, a lower mass bound of the gluino mass of $500\GeV$ holds 
(see Ref.~\cite{Dreiner:2012gx} for a detailed discussion on how the LHC bounds change). 

Theoretical predictions for gluino pair production up to next-to-leading-order
(NLO) in QCD have been obtained in Ref.~\cite{Beenakker:1996ch}.
The hadronic leading order (LO) and NLO cross sections can be evaluated 
numerically using the program \texttt{Prospino}~\cite{Beenakker:1996ed}. 
As an improvement beyond NLO, the threshold enhanced logarithms 
have been resummed to next-to-leading-logarithmic (NLL) accuracy~\cite{Kulesza:2008jb,Kulesza:2009kq,Beenakker:2009ha},
implying corrections of about $2-35\%$ in comparison to the NLO cross section 
which depend on the gluino mass and the chosen parton distributions (PDFs).
In Ref.~\cite{Kauth:2011vg}, threshold effects at NLO in QCD 
due to remnants of the $1S$ resonance of gluino bound states are discussed 
leading to an enhancement of the complete NLO threshold cross section 
of $7-9\%$ compared to the fixed order predictions.
Recently, the combined NLL resummation of threshold logarithms and the Coulomb
corrections for gluino pair production has been studied in Ref.~\cite{Falgari:2012hx}
and phenomenological predictions for cross sections at the LHC have been 
summarized in Ref.~\cite{Kramer:2012bx}.
The inclusive cross section for squark-antisquark pair production has been subject to 
similar improvements in the past~\cite{Kulesza:2009kq,Langenfeld:2009eg} 
and presently, the corresponding predictions beyond NLO are exact to 
next-to-next-to-leading-logarithmic (NNLL) accuracy~\cite{Beenakker:2011sf}.
In contrast, the available results for gluino pair production are still
limited to NLL accuracy, only.

In this article, we improve the available QCD predictions 
for gluino pair production to NNLL accuracy, putting it on par 
with the case of squark-antisquark pair production.
To that end, we compute the missing hard matching coefficients 
at NLO near threshold.
With our new results, we are able to provide QCD predictions 
for the total hadronic cross sections at approximately next-to-next-to-leading order (NNLO).
These corrections lead to a further increase of the cross section of the order
of $10\%$ in comparison to the NLO results.
As all searches for SUSY particles so far have resulted in exclusion limits only, 
a precise knowledge of the gluino pair production cross section
in the threshold region is of special interest, 
because the size of the expected rates has 
a direct impact on the excluded mass range for gluinos.

The article is organized as follows.
In Sec.~\ref{sec:theo}, we recall the basic ingredients of the 
hadronic and partonic production cross sections.
In Sec.~\ref{sec:partonic}, we review the formalism of threshold resummation in Mellin space
and then proceed to extract the color-decomposed NLO cross 
section at the threshold from known results for gluino-bound state
production given in Ref.~\cite{Kauth:2011vg}. 
Verifying a general result of Ref.~\cite{Beneke:2009ye}, we then
calculate the color-decomposed NNLO cross section in the threshold limit.  
Finally, we resum the cross section to NNLL accuracy, matched onto 
the approximated NNLO result.
We check our analytic formulas by extracting the
color-summed one-loop matching constants from \texttt{Prospino} via an appropriate 
fit in the threshold region.
The hadronic production cross section is discussed in Sec.~\ref{sec:hadronic}. 
The appendices contain useful analytical expressions for certain scalar $n$-point integrals 
and the expansion coefficients of the general resummation formula.

\section{Theoretical Setup}
\label{sec:theo}
\begin{figure}
\centering
 \scalebox{0.44}{\includegraphics{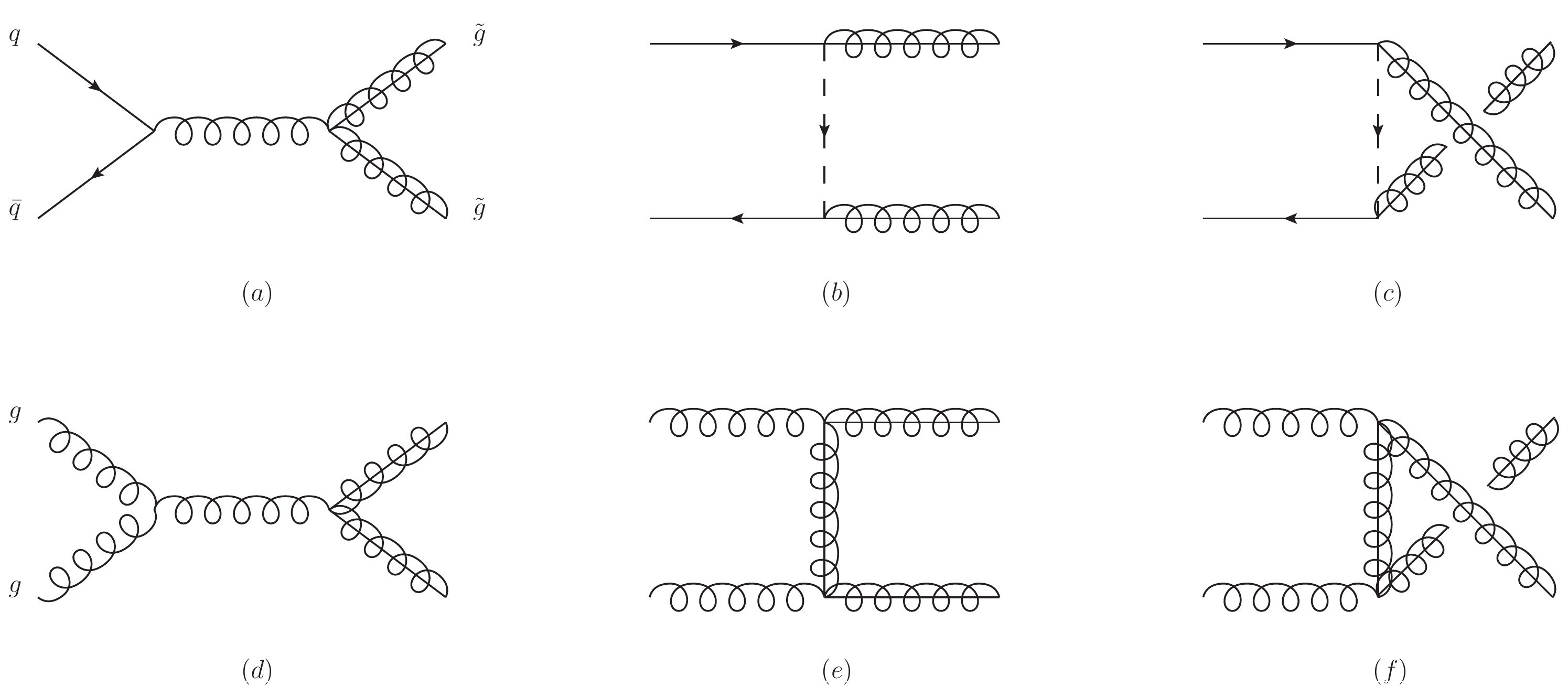}}
\vspace*{3mm}
\caption{Feynman diagrams for the production of a gluino pair \gluinopair~via $\qqbar$ annihilation (diagrams (a) - (c)) 
        and $gg$ annihilation (diagrams (d) - (f)) at LO.}
\label{fig:LOdiagrams}
\end{figure}

We study the hadro-production of gluino pairs at the LHC (i.e. the reaction $\pptogluinopair$) with its partonic sub processes
\begin{eqnarray}
  \label{eq:procgg}
  g g  &\to& \widetilde{g} \widetilde{g}
  \, ,\\
  \label{eq:procqq}
  q\bar{q} &\to& \widetilde{g} \widetilde{g}, \enspace\enspace q = d,u,s,c,b \enspace
  \, ,
\end{eqnarray}
with the relevant LO Feynman diagrams shown in Fig.~\ref{fig:LOdiagrams}.

We focus on the inclusive hadronic cross section of hadro-production 
of gluino pairs, $\sigma_{p p \rightarrow \gluinopair X}$, which is a function
of the hadronic cms energy $\sqrt{s}$, 
the gluino mass $\mg$, the squark mass $\mq$ 
(assuming mass degeneracy among the squark flavors), 
and the renormalization and factorization scales, $\mu_r$ and $\mu_f$, respectively.
In the standard factorization approach of perturbative QCD, it reads
\begin{eqnarray}
  \label{eq:totalcrs}
  \sigma_{pp/p\bar{p} \to \gluinopair X}(s,\mg^2,\mq^2,\mufs,\murs) &=&
  \sum\limits_{i,j = q,{\bar{q}},g} \,\,\,
  \int\limits_{4\mg^2}^{s }\,
  d {\shat} \,\, L_{ij}(\shat, s, \mufs)\,\,
  \hat{\sigma}_{ij \to \gluinopair} ({\shat},\mg^2,\mq^2,\mufs,\murs)
  \, ,
\end{eqnarray}
where the parton luminosities $L_{ij}$ are given as convolutions
of the PDFs $f_{i/p}$ defined through
\begin{eqnarray}
  \label{eq:partonlumi}
  L_{ij}(\shat, s, \mufs) &=&
  {\frac{1}{s}} \int\limits_{\shat}^s
  {\frac{dz}{z}} f_{i/p}\left(\mufs,{\frac{z}{s}}\right)
  f_{j/p}\left(\mufs,{\frac{\shat}{z}}\right)
  \, .
\end{eqnarray}
Here, $\shat$ denotes the partonic cms energy.
As will be discussed below, the treatment of soft-gluon
exchange in the final-state gluino pair requires the knowledge
of the color-decomposed partonic cross sections $\hat{\sigma}_{ij,\,{\bf I}}$.
Setting $\mu_f=\mu_r=\mu$, the latter is commonly expressed by dimensionless 
scaling functions $f^{(kl)}_{ij\,{\bf I}}$
in a power series in the strong coupling constant $\alpha_s=\alpha_s(\mu)$,
\begin{eqnarray}
  \label{eq:definef}
  \hat{\sigma}_{ij,\,{\bf I}} &=& \frac{\as^2}{\mg^2}
  \Biggl[
    f^{(00)}_{ij,\,{\bf I}} + 4\pi\as \Bigl(f^{(10)}_{ij,\,{\bf I}} 
    + f^{(11)}_{ij,\,{\bf I}}L_\mu\Bigr)
    + (4\pi\as)^2\Bigl(f^{(20)}_{ij,\,{\bf I}} + f^{(21)}_{ij,\,{\bf I}}L_\mu 
    + f^{(22)}_{ij,\,{\bf I}}L_\mu^2 \Bigr)
    + \ord\left(\alpha_s^3\right)\Biggr]
  \, ,
  \quad
\end{eqnarray}
with $L_\mu = \ln({\mu^2}/{\mg^2})$.
We use the capital index ${\bf I}$ to label the admissible $SU(3)_{\mathrm{color}}$
representations of the scattering reactions~(\ref{eq:procgg}) and (\ref{eq:procqq}). 
The decomposition of the gluon-fusion channel~(\ref{eq:procgg}) into irreducible color
representations is given by
\begin{eqnarray}
  \label{eq:colgg}
 \mathbf{8}\times\mathbf{8} &=& \mathbf{1}_s + \mathbf{8}_s + \mathbf{8}_a +
  \mathbf{10} + \mathbf{\overline{10}} + \mathbf{27}_s
  \, ,
\end{eqnarray}
and a suitable basis in terms of the generators of the 
adjoint representation of the $SU(3)_{\mathrm{color}}$ can be found
in~\cite{Bartels:1993ih} (see also~\cite{Kulesza:2009kq}).
Likewise, for the quark-antiquark channel~(\ref{eq:procqq}), we use the color basis
\begin{eqnarray}
  \label{eq:colqq}  
 \mathbf{3}\times\mathbf{{\bar 3}} &=& \mathbf{1}_s + \mathbf{8}_s + \mathbf{8}_a 
  \, .
\end{eqnarray}
The partonic cross sections in Eq.~(\ref{eq:totalcrs}) are recovered 
after summation over all color structures,
\begin{eqnarray}
\label{eq:colorsum}
\hat{\sigma}_{ij \to \gluinopair} &=& 
\sum\limits_{{\bf I}}\, \hat{\sigma}_{ij,\,{\bf I}} 
\, ,
\end{eqnarray}
and, similarly, for the scaling functions in Eq.~(\ref{eq:definef}).
The (color-summed) scaling functions at LO are given by~\cite{Beenakker:1996ch} 
\begin{eqnarray} 
\label{eq:fgg00}
f^{(00)}_{gg} &=&
\frac{\pi}{4}\rho\Biggl[-3\beta\Biggl(1 + \frac{17}{16}\*\rho\Biggr)
  + \frac{9}{4}\Biggl(-1 - \rho + \frac{1}{4}\*\rho^2\Biggr)L_1\Biggr]
=\, \frac{27}{64}\,\pi \beta\,+\mathcal{O}(\beta^3)\,,
\\[2mm]
\label{eq:fqq00}
f^{(00)}_{\qqbar} &=&
\frac{\pi}{9}\,\*\beta\*\rho\*\big(2+\rho\big)
+\frac{\pi}{6}\,\*\rho\*\Big[-\beta\*\big(2+q\*\rho\big) 
  +\big(1 + \tfrac{1}{4}\*q^2\*\rho\big)\*\rho\*L_2\Big]
\\
&& \quad    +\frac{\pi}{27}\,\*\rho\Biggl[16\*\beta\*
  \Biggl(\frac{2 - 2\*q + q^2\*\rho}{4 - 4\*q+q^2\*\rho}\Biggr)
  -\Biggl(\frac{1 + 8\*q - 4\*q^2\*\rho}{2-q\*\rho}\Biggr)\*\rho\*L_2\Biggr]
\non\\[2mm]
&=& \frac{\pi}{3} \*\Biggl(\frac{1-r}{1+r}\Biggr)^2 \*\beta  
- \frac{4\*\pi}{81}\,\*\frac{(4 + 9\*r - 77\* r^2 + 27 \*r^3 + 9 \*r^4)}{(1+r)^4}\,\*\beta^3
+ \mathcal{O}(\beta^5)
\, .
\non
\end{eqnarray}
with the abbreviations
\begin{eqnarray}
\label{eq:def-variables}
r = \frac{\mq^2}{\mg^2},\enspace
q = 1 - r,\enspace
\rho = \frac{4\mg^2}{\shat},\enspace \beta = \sqrt{1 - \rho},\enspace
L_1 = \ln\left(\frac{1 - \beta}{1 + \beta}\right) ,\enspace
L_2 = \ln\left(\frac{2-q\*\rho - 2\*\beta}{2-q\*\rho + 2\*\beta}\right). 
\end{eqnarray}
The threshold expansion of the LO hard function for gluon fusion 
depends only on the dimensionless variable $\beta$, which is zero 
at the threshold $\shat = 4\mg^2$.
The expanded LO function for quark-antiquark annihilation depends on 
$\beta$ and on the ratio $r$ of the squared squark to gluino masses. 
For $r=1$, the linear term vanishes. 
Therefore, small mass differences between the gluino and the squark mass lead
to a suppression of the \qqbar~channel. 
The $gq$~channel on the other hand is absent at tree level.
Its NLO contribution at threshold is of the order $\beta^3 \ln(\beta)$ 
and thus strongly suppressed compared to the $gg$ and $\qqbar$ channels. 
We will therefore ignore its contribution to gluino production
in the following discussion. 
However, we include its NLO contribution to the total hadronic cross section.

For the color decomposition of Eq.~(\ref{eq:colgg}) 
we find in agreement with Ref.~\cite{Kulesza:2009kq} the following Born
scaling functions for the reaction~(\ref{eq:procgg}),
\begin{eqnarray}
\label{eq:gg}
f_{gg,\,{\bf I}_s}^{(00)} &=& 
- N_{\bf I}\*\frac{9\pi}{128}\,\*\rho\*\Biggl[\beta\bigl(1+\rho\bigr)
+\frac{1}{2}\*\bigl(2+2\*\rho-\rho^2\bigr)\*L_1
\Biggr] 
\,=\, 
N_{\bf I}\*\frac{9\pi}{128}\*\,\beta +\mathcal{O}(\beta^3)
\, ,
\\[2mm]
f_{gg,\,{\bf 8}_a}^{(00)} &=&
- \frac{3\pi}{128}\*\,\rho\*\Biggl[2\*\beta\*\bigl(7+8\*\rho\bigr)
+3\*\bigl(2 + 2\*\rho + \rho^2\bigr)\*L_1
\Biggr] = \mathcal{O}(\beta^3),
\\[2mm]
f_{gg,\,{\bf 10+\overline{10}}}^{(00)} &=&\, 0
\, ,
\end{eqnarray}
with $L_1$ given in Eq.~(\ref{eq:def-variables}).
Note, that in the threshold limit, only symmetric color representations in
Eq.~(\ref{eq:gg}) contribute, for which we define the normalization factor,
\begin{eqnarray}
\label{eq:NI}
N_{\bf I} \,=& \lbrace 1,2,3\rbrace
\hspace*{17mm} 
& \text{for}\quad {\bf I}=\lbrace {\bf 1,8,27} \rbrace 
\, .
\end{eqnarray}

For $\qqbar$ annihilation in Eq.~(\ref{eq:procqq}), we obtain the LO scaling functions 
in the color decomposition of Eq.~(\ref{eq:colqq}) as
\begin{eqnarray}
\label{eq:qqbarsinglet}
 f_{\qqbar,\,\mathbf{1}_S}^{(00)} &=& \frac{\pi}{27}\,\*
        \rho\bigl(2 - 2\*q + q^2\*\rho\bigr)\*\Biggl[\frac{2\beta}{4 - 4\*q +q^2\*\rho}
        +\frac{1}{2}\frac{1}{2-q\*\rho}\*\rho\*L_2\Biggr]
      = \frac{16}{81}\*\pi\*\frac{(1+r^2)}{(1+r)^4}\*\beta^3
      +\mathcal{O}(\beta^5)
\, ,\\[2mm]
\label{eq:qqbaroctetts}
f_{\qqbar,\,\mathbf{8}_S}^{(00)} &=& \frac{\pi}{27}\,\*
      \rho\bigl(2 - 2\*q + q^2\*\rho\bigr)\*\Biggl[\frac{5\beta}{4 - 4\*q +q^2\*\rho}
        +\frac{5}{4}\frac{1}{2-q\*\rho}\*\rho\*L_2\Biggr]
      = \frac{40}{81}\*\pi\*\frac{(1+r^2)}{(1+r)^4}\*\beta^3 
      + \mathcal{O}(\beta^5)
\, ,\\[2mm]
\label{eq:qqbaroctetta}
f_{\qqbar,\,\mathbf{8}_A}^{(00)} &=&
   \frac{\pi}{9}\,\*\beta\rho\*\bigl(2 + \rho \bigr)
  + \frac{\pi}{6}\rho\Bigl[-\beta\bigl(2+q\*\rho\bigr) + \bigl(1 +
    \tfrac{1}{4}q^2\*\rho\bigr)\rho L_2\Bigr]
\\[2mm]
  &&\hspace*{10mm}
  +\frac{\pi}{27}\,\*
      \rho\*\Biggl[\frac{9\beta(2 - 2\*q + q^2\*\rho)}{4 - 4\*q +q^2\*\rho}
        -\frac{9}{4}\frac{2 + 2\*q - q^2\*\rho}{2-q\*\rho}\*\rho\*L_2\Biggr]
\nonumber\\[2mm]
&=&\frac{\pi}{3} \*\Biggl(\frac{1-r}{1+r}\Biggr)^2 \*\beta 
-\frac{4}{9}\*\pi\*\frac{(1-r)^2(2+5\*r+r^2)}{(1+r)^4}\*\beta^3
+ \mathcal{O}(\beta^5)
\, ,
\nonumber
\end{eqnarray}
with $L_2$ given in Eq.~(\ref{eq:def-variables}).
Note that at threshold, only the antisymmetric octet representation of the 
$q\bar{q}$ channel contributes
if the gluino and the squark masses are different.
If the gluino and the squarks have equal masses, 
the antisymmetric octet scaling function is vanishing up to $\ord(\beta^4)$, 
see Eq.~(\ref{eq:qqbaroctetta}), and the symmetric singlet and octet scaling function
contribute with the ratio $2 \,: \,5$ at the production threshold.
If $r \neq 1$, the gluino pairs are produced in an $S$-wave, otherwise in a $P$-wave
in that channel.

\section{Higher order partonic cross sections at the threshold}
\label{sec:partonic}

At higher orders in QCD, the cross sections develop large threshold logarithms of the type $\ln(\beta)$ in 
the region $\hat s\approx 4\mg^2$, which can be resummed systematically 
to all orders in perturbation theory. Here, we make use of techniques described 
in~\cite{Sterman:1986aj,Catani:1989ne,Catani:1990rp,Contopanagos:1996nh,Catani:1996yz,Kidonakis:1997gm,Moch:2005ba}.
The resummation is performed in Mellin space after introducing
moments $N$ with respect to the variable $\rho = 4\mg^2/\shat\,$
of momentum space,
\begin{eqnarray}
  \label{eq:mellindef}
  \hat{\sigma}(N,\mg^2) &=&
  \int\limits_{0}^{1}\,d\rho\, \rho^{N-1}\,
  \hat{\sigma}(\shat,\mg^2)\, .
\end{eqnarray}
As the threshold limit $\beta\to 0$ corresponds to $N\to\infty$, all terms proportional
to powers of $1/N$ will be discarded.
The general resummation formula reads
\begin{eqnarray}
  \label{eq:sigmaNres}
 \hat{\sigma}_{ij,\, {\bf I}}(N,\mg^2)       
  &=&
  \hat{\sigma}^{B}_{ij,\, {\bf I}}(N,\mg^2)\,
  g^0_{ij,\, {\bf I}}(N+1,\mg^2) \, \exp \Big[ G_{ij,\,{\bf I}}(N+1) \Big] +
  {\cal O}(N^{-1}\ln^n N) \, ,
\end{eqnarray}
where we have suppressed all dependence on the renormalization and
factorization scale, $\mur$ and $\muf$. 
The subscripts $ij$ denote the production channel, where we consider $ij=gg,\qqbar$.
The exponent $G_{ij,\, {\bf I}}$ contains all large threshold logarithms $\ln^k N$ in Mellin-$N$ space, 
and the resummed cross section, as indicated in Eq.~(\ref{eq:sigmaNres}), 
is accurate up to terms which vanish as a power for large Mellin-$N$. 
To NNLL accuracy, $G_{ij,\, {\bf I}}$ is commonly expanded as
\begin{eqnarray}
  \label{eq:GNexp}
  G_{ij,\, {\bf I}}(N) =
  \ln(N) \cdot g^1_{ij}(\lambda)  +  g^2_{ij,\, {\bf I}}(\lambda)  +
  a_s\, g^3_{ij,\, {\bf I}}(\lambda)  + \dots\, ,
\end{eqnarray}
where $\lambda = a_s\,\beta_0\, \ln N$ and we abbreviate $a_s = {\alpha_s }/{(4 \pi)}$.
The functions $g^k_{ij,\, {\bf I}}$ 
are derived from the double integral over a set of anomalous dimensions 
(see e.g.,~\cite{Moch:2005ba,Moch:2008qy,Beneke:2009rj}),
\begin{eqnarray}
  \label{eq:GN}
  G_{ij,\,{\bf I}}(N) &=& 
  \int_0^1 dz\;\frac{z^{N-1}-1}{1-z}\,
  \Bigg{\lbrace}\int_{\mu_f^2}^{4\mg^2(1-z)^2}                        
  \frac{dq^2}{q^2}\;\Big{(} A_i\big{(}\as(q^2)\big{)}
  + A_j\big{(}\as(q^2)\big{)}\Big{)}
\\ 
&& \qquad\qquad\qquad\qquad\qquad  
  + D_{ij,\,{\bf I}}\big{(}\as(4\mg^2(1-z)^2)\big{)} \Bigg{\rbrace}
  \, .
\nonumber
\end{eqnarray}
Here, the cusp anomalous-dimension $A_{i}$ refers to initial-state
collinear gluon radiation, while any large-angle soft gluon 
radiation is contained in the function $D_{ij,\,{\bf I}}$, which splits into the functions 
\begin{eqnarray}
D_{ij,\,{\bf I}}(\as) &=& \frac 12\,\Big{(}D_i(\as) +\, D_j(\as)\Big{)}\,
+\, D_{\widetilde g \widetilde g,\,{\bf I}}(\as)
\, ,
\end{eqnarray}
for initial- and final-state radiation, where $D_i$ can be taken from 
threshold resummation for the Drell-Yan process or for Higgs production 
in gluon fusion.
The perturbative expansion for the anomalous dimensions reads
\begin{eqnarray}
\label{eq:fctexpand}
A_i(\as) &=& \sum_l \left(\frac{\as}{4\pi}\right)^l\,A_i^{(l)} \equiv \sum_l a_s^l\,A_i^{(l)}
\, ,
\end{eqnarray}
(same for $D_i(\as)$ etc.) 
and the expansion coefficients $A_i^{(l)}$ and $D_i^{(l)}$ are both known to third order 
in $a_s$ from Refs.~\cite{Moch:2004pa,Vogt:2004mw} and~\cite{Moch:2005ky,Laenen:2005uz},
respectively.
The function $D_{\widetilde g \widetilde g,\,{\bf I}}^{(l)}$ 
due to soft gluon emission in the final state depends 
on the $SU(3)_{\mathrm{color}}$ representation of the final-state gluino pair 
and results up to second order in $a_s$ are given 
in Ref.~\cite{Beneke:2009rj} for heavy final states in arbitrary 
color representations\footnote{
Ref.~\cite{Beneke:2009rj} uses a different notation: The coefficients are
denoted by $D_{HH'}^{(n)R_\alpha}$, where $n=l-1$. }
(see also~\cite{Becher:2009kw}).
This suffices to compute the functions $g^{(l)}_{ii, {\bf I}}$ in Eq.~(\ref{eq:GNexp})
to NNLL accuracy, even with the dependence on the $\mur$ and $\muf$ separated
(for the computation see, e.g.,~Refs.~\cite{Vogt:2000ci,Moch:2005ba}).
The explicit expression for $g^{(1)}_{ii}$ can be read off from Eq.~(A.5) of 
Ref.~\cite{Moch:2008qy}, 
for $g^{(2)}_{ii}$ from Eq.~(A.7) and for $g^3_{ii, {\bf I}}$ from Eq.~(A.9) 
of that reference with the replacements 
$A^{(l)}_q \to A^{(l)}_i/\beta^{(l)}_0$, 
$D^{(l)}_q \to D^{(l)}_i/\beta^{(l)}_0$, 
$D_{Q\bar{Q}} \to - D^{(l)}_{gg,\,{\bf I}}/\beta^{(l)}_0$ 
(note the sign convention for $D_{Q\bar{Q}}\:\!$ in~\cite{Moch:2008qy}), and 
$\beta_l \to \beta_l/\beta^{(l+1)}_0$, 
where $\beta_l$ denote the well-known QCD beta-function coefficients in the
normalization~(\ref{eq:fctexpand}). 

As a last remaining step in achieving resummed predictions to NNLL accuracy in QCD, 
one has to extract the process-dependent matching constants $g^0_{ij,\,{\bf I}}$ in Eq.~(\ref{eq:sigmaNres}).
These consist of the hard coefficients $g^0_{ij,\,{\bf I}}(\as)$ multiplied by 
Coulomb coefficients $g^{0,\,C}_{ij,\,{\bf I}}(\as,N)$, which also account for 
the interference of Coulomb exchange with hard contributions and soft radiation.  
A perturbative expansion in analogy to Eq.~(\ref{eq:fctexpand}) yields
\begin{eqnarray}
\label{eq:g0fac}
g^{(0)}_{ij,\, {\bf I}}(N,\mg^2)
&=& g^{(0)}_{ij,\,{\bf I}}(\as)\,g^{(0),\,C}_{ij,\,{\bf I}}(\as,N)
\\
&=& 1 + a_s\left(g^{(0)\,(1)}_{ij,\,{\bf I}} + g^{(0),\,C\,(1)}_{ij,\,{\bf I}}(N)\right)
\nonumber\\
& & 
\hspace*{3mm}
+ a_s^2\left(g^{(0)\,(2)}_{ij,\,{\bf I}} + g^{(0),\,C\,(2)}_{ij,\,{\bf I}}(N)
                   + g^{(0)\,(1)}_{ij,\,{\bf I}}\,g^{(0),\,C\,(1)}_{ij,\,{\bf I}}(N)\right) 
  + \ord\left(\alpha_s^3\right)
\nonumber
\, .
\end{eqnarray}
This factorized form is already known from studies of the QCD hadro-production
of heavy quarks (see also Ref.~\cite{Kawamura:2012cr})
and allows for a separate treatment of the resummation of threshold logarithms 
$\as^n\ln^m\beta$ (hard, $m\leq 2n$) and the terms proportional to $\as^n\beta^{-m}\ln^l\beta$ 
(Coulomb, $m\leq n$).
Note, that the matching constant $g^{(0)}_{ij,\,{\bf I}}(\as)$ in the first case does not depend on the Mellin 
moment $N$, whereas in the second case $g^{(0),\,C}_{ij,\,{\bf I}}(\as,N)$ does.
In the following, we will focus on the computation of the one-loop hard matching coefficients 
which is the main new result of the present paper and which allows for the
extraction of the expansion coefficients of $g^{(0)}_{ij,\,{\bf I}}$ in Eq.~(\ref{eq:g0fac}) 
to NNLL accuracy. All explicit expressions are given in App.~\ref{sec:gi}.

Before doing so, we briefly like to comment on the resummation of Coulomb corrections, 
which accounts for the bound-state effects in the gluino pair \cite{Hagiwara:2009hq} and 
which exploits an effective description of QCD in the non-relativistic regime.
To leading power, it is long known that the so-called 
Sommerfeld factor $\Delta^C$ sums the pure Coulomb corrections in momentum space ($\beta$)
corresponding to ladder diagrams~\cite{Fadin:1990wx}.
One has~\cite{Kulesza:2009kq},
\begin{eqnarray}
  \label{eq:DeltaC}
\Delta^C &=& 
\Delta^C\left(\frac{\pi\,\as}{\beta}\,D_{\bf I}\right)\,,
\qquad \Delta^C(x)= \frac{x}{{\rm exp}(x)-1}\,,
\end{eqnarray}
where we have introduced the quantity $D_{\bf I} = C_{\bf I}/2-C_A$
as a function of $C_A=3$ 
and the quadratic Casimir operators $C_{\bf I}$ of the final-state $SU(3)_{\mathrm{color}}$ representation.
For initial-state gluons these take the numerical values 
\begin{eqnarray}
\label{eq:CI}
C_{\bf I} \,=& \lbrace 0,3,8 \rbrace
\hspace*{17mm} 
& \text{for}\quad {\bf I}=\lbrace {\bf 1,8,27} \rbrace 
\, ,
\\
\label{eq:DI}
D_{\bf I} \,=& \lbrace -3,-3/2,\,1 \rbrace
\hspace*{5mm} 
& \text{for}\quad {\bf I}=\lbrace {\bf 1,8,27} \rbrace
\, ,
\end{eqnarray}
depending on the gluino pair being in the ${\bf I}=\lbrace {\bf 1,8,27} \rbrace$-representation.
In the $\qqbar$~channel, only the anti-symmetric octet contributes
at first order in the threshold expansion and we always set $C_{\bf I}=C_A=3$, 
thus it follows $D_{\bf I} = -3/2$ from Eq.~(\ref{eq:DI})\footnote{
The notations for $D_{\bf I}$ vary in the literature:
$\kappa_{ij\to\widetilde g\widetilde g}$ in Ref.~\cite{Kulesza:2009kq},
$C^{[R]}$ in Ref.~\cite{Kauth:2011vg} and $D_{R_\alpha}$ in Ref.~\cite{Beneke:2009ye} 
(where it is explicitly given for top-quark production, 
which differs from the color configurations of gluino production).}.

A formal expansion of Eq.~(\ref{eq:DeltaC}) in $\as$ reproduces the NLO and NNLO pure 
Coulomb terms to leading power.
However, the expansion does not converge due to the high inverse powers of $\beta$ 
close to the threshold 
and those singular terms even cause a fixed-order expansion of
Eq.~(\ref{eq:mellindef}) beyond NNLO to be ill-defined. 
In the context of hadronic heavy-quark production 
this has motivated detailed studies of the phenomenological effects of
Coulomb resummation~\cite{Beneke:2011mq}. 
Methods and results for the combined resummation of threshold logarithms 
and the Coulomb corrections for heavy quarks 
have also been presented in \cite{Beneke:2009rj,Beneke:2010da,Falgari:2012hx}.
The effect of Coulomb resummation for the total cross section is small, e.g.
$\ord(1\%)$ for the related case of heavy-quark hadro production.

\bigskip

Let us now turn to the calculation of the necessary one-loop hard matching coefficients.
To that end, recall that the NLO scaling functions 
$f^{(10)}_{gg}$ and $f^{(10)}_{\qqbar}$ near threshold 
can be written in a factorized form with respect to the Born contributions 
as~\cite{Beenakker:1996ch}
\begin{eqnarray}
\label{eq:fgg10}
f^{(10)}_{gg} &=& \frac{f^{(00)}_{gg}}{4\pi^2}\Biggl(
6 \*\ln^2\big(8\*\beta^2\big) - 29\*\ln\big(8\*\beta^2\big)
+ \frac{\pi^2}{4\beta} + C_{1}^{gg} \Biggr)
\, ,\\[2mm]
\label{eq:fqq10}
f^{(10)}_{\qqbar} &=& \frac{f^{(00)}_{\qqbar}}{4\pi^2}
\Biggl(\frac{8}{3}\*\ln^2\big(8\*\beta^2\big) 
- \frac{41}{3}\*\ln\big(8\*\beta^2\big) 
+ \frac{3\*\pi^2}{4\*\beta} + C_{1}^{\qqbar}\Biggr)
\, ,
\end{eqnarray}
where $C_{1}^{gg}$ and $C_{1}^{\qqbar}$ define the hard one-loop constants 
to be determined from matching to a fixed order NLO calculation near the threshold.
For the former quantity, we actually need the individual components $C_{1,\,{\bf I}}^{gg}$ with respect
to the final-state color configuration.
Therefore, we decompose Eq.~(\ref{eq:fgg10}) as 
\begin{eqnarray}
\label{eq:fgg10I}
f^{(10)}_{gg,\,{\bf I}} &=& \frac{f^{(00)}_{gg,\,{\bf I}}}{4\pi^2} 
\Biggl(
6 \*\ln^2\big(8\*\beta^2\big) - (24+C_{\bf I})\*\ln\big(8\*\beta^2\big)
- \frac{\pi^2}{2\*\beta}\*D_{\bf I} + C_{1,\,{\bf I}}^{gg} \Biggr)
\, ,\quad
\end{eqnarray}
where $D_{\bf I}$ and $C_{\bf I}$ are defined in Eqs.~(\ref{eq:CI}) and ~(\ref{eq:DI}).
Note, that only the color-symmetric states ${\bf 1,8,27}$ contribute to the gluon-fusion channel, 
cf. Eq.~(\ref{eq:colgg}), whereas the anti-symmetric octet scaling function is
suppressed in the threshold limit and, therefore, neglected. 
For quark-antiquark annihilation, 
only the anti-symmetric octet channel is considered, see Eq.~(\ref{eq:colqq}).
Summation over all color configurations defines $C_1^{gg}$ in Eq.~(\ref{eq:fgg10})
as
\begin{eqnarray}
\label{eq:Cgg1}
C_1^{gg} &=&\frac{\sum_{\bf I} N_{\bf I} C_{1,\,{\bf I}}^{gg}}
{\sum_{\bf I} N_{\bf I}}
\, ,
\end{eqnarray}
with $N_{\bf I}$ given in Eq.~(\ref{eq:NI}).

The analytic expressions for the one-loop matching constants 
$C_{1,\,{\bf I}}^{gg}$ and $C_{1}^{\qqbar}=C_{1,\,{\bf 8}_a}^{\qqbar}$
can be extracted from Ref.~\cite{Kauth:2011vg}, where the authors
studied the QCD effects for a gluino bound state $T$ with an invariant mass $M$.
At the production threshold for gluino pairs, the differential cross section reads 
(cf. Eq.~(5) in~\cite{Kauth:2011vg})
\begin{eqnarray}
\label{eq:boundstate}
M\,\frac{d \sigma_{ij\to T,\,{\bf I}} }{dM} (\hat s,M^2,\mu_f^2,\mu_r^2)&=& 
{\mathcal F}_{ij\to T,\,{\bf I}}(\hat s,M^2,\mu_f^2,\mu_r^2)\,
\frac{1}{\mg^2}\,
\text{Im}\lbrace {\mathcal G}_{\bf I}(0,M-2\mg + i\Gamma_{\tilde g} \rbrace
\, ,
\end{eqnarray}
where ${\mathcal F}_{ij\to T,\,{\bf I}}\,$ denotes the hard scattering kernel, 
${\mathcal G}_{\bf I}$ the Green's function of a non-relativistic Schr{\"o}dinger 
equation, which accounts for the binding effects, and $\Gamma_{\tilde g}$ the gluino decay width.
Suppressing higher powers of $\beta$, the hard function can be factorized as
(cf. Eq.~(19) in~\cite{Kauth:2011vg})
\begin{eqnarray}
\label{eq:hardfunc}
{\mathcal F}_{ij\to T,\,{\bf I}} &=& 
{\mathcal F}_{ij\to T,\,{\bf I}}^{\rm Born}\,
\left(1+\,\frac{\as^{\rm \overline{DR}}(\mu_r)}{\pi}\,\overline{\mathcal V}_{ij,\,{\bf I}}\right) 
\left[\delta(1-z) 
  +\,\frac{\as^{\rm \overline{DR}}(\mu_r)}{\pi}
  \,\overline{\mathcal R}_{ij,\,{\bf I}}(z)
  \right]
\, ,
\end{eqnarray} 
where $z=M^2/\hat s$. 
$\overline{\mathcal V}_{ij,\,{\bf I}}$ denotes the infrared-finite parts of the 
ultraviolet-regularized virtual corrections
and $\overline{\mathcal R}_{ij,\,{\bf I}}$ the real corrections. 
In the threshold limit, these quantities are unaffected by the dynamics of 
the bound state formation. Thus, the explicit expressions can be taken over
for the calculation of the gluino pair production cross section
in the threshold region, where the (imaginary part of the) 
Green's function in Eq.~(\ref{eq:boundstate}) is set to one.
The difference between the $2\to 1$ and $2\to 2$ kinematics is encoded in the
Born term ${\mathcal F}_{ij\to T,\,{\bf I}}^{\rm Born}\,$.
In order to obtain the NLO hard kernels for gluino pair production, we simply 
have to replace the latter by our LO functions (\ref{eq:gg}) and (\ref{eq:qqbaroctetta}).
Setting the binding energy to zero, we further replace $M$ by $2\mg$, 
and thus $z$ by $\rho$ in Eq.~(\ref{eq:hardfunc}).

In the related case of the QCD corrections to hadronic top-quark pair production, 
this procedure has been discussed in Ref.~\cite{Langenfeld:2009wd}, 
showing that the required NLO matching of the inclusive cross section to NNLL
accuracy near threshold including the decomposition for color-singlet and color-octet states 
can be performed with the help of the NLO QCD corrections to
hadro-production of quarkonium~\cite{Petrelli:1997ge} (see also Refs.~\cite{Hagiwara:2008df,Kiyo:2008bv}).

In the full MSSM, the number of both the quark and squark flavors, 
that enter the virtual NLO contributions, is given by $n_f=6$ 
and we set all squark masses equal. 
The one-loop matching coefficients $C_{1,\,{\bf I}}^{ij}$ depend on the 
chosen regularization scheme and we find in dimensional reduction
$\overline{\rm DR}$ adopted in Ref.~\cite{Kauth:2011vg},
\begin{eqnarray}
\label{eq:C1ggI}
C_{1,\,{\bf I}}^{gg\,\overline{\rm DR}} &=& 
 C_{\bf I}\*\left(4 + \ln(2) -\frac{\pi^2}{8}\right) 
        + \frac{71}{2}\, - 6\,\*\ln^2(2) -\pi^2
        + \frac{1}{3}\,\*L_{t\widetilde g} + \frac{n_f}{6}\,\*\ln(r)
        + \frac{n_f}{18}\,\*A_{\bf I}^{gg}(r)
\, ,\\
\label{eq:C1qq8}
C_{1,\,{\bf 8}_a}^{\qqbar\,\overline{\rm DR}} &=& 
       n_f\*\left(\ln(2)- \frac{5}{9}\right) 
          + \frac{89}{3}
          - \frac{8}{3}\,\*\ln^2(2) 
          - \frac{43}{36}\,\*\pi^2 + A^{\qqbar}_{{\bf 8}_a}(r)
\,,
\end{eqnarray}
where we have defined $L_{t\widetilde g} = \ln(m_t^2/\mg^2)$, as well as
\begin{eqnarray}
\label{eq:Agg}
A_{\bf I}^{gg}(r) &=& 
  - 9\,\*\Big{(}b_1(r) - b_4(r) + 2\,\*b'_1(r)\Big{)}\,\*(1 - r) 
  + 2\,\*\Big{(}9\*r - (C_{\bf I}+1)\Big{)}\,\*c_5(r)
\, ,
\end{eqnarray}
while $A^{\qqbar}_{{\bf 8}_a}(r)$ is given in Eq.~(31) of Ref.~\cite{Kauth:2011vg}.
The results involve functions $a_1(r)$, $b_{i}(r)$, $b'_{i}(r)$ and $c_{i}(r)$ 
which have been defined as certain limits  of scalar one-, two-, and three-point 
integrals in the Appendix of Ref.~\cite{Kauth:2011vg}.
For the convenience of the reader, we give the explicit analytic expressions in 
App.~\ref{sec:npoint}. 

The quantity $A^{\qqbar}_{{\bf 8}_a}(r)$ in Eq.~(\ref{eq:C1qq8}) diverges in the limit $r\to 1$,
but multiplication with the Born cross section gives a finite result 
for the NLO scaling function in the threshold approximation (which is actually zero).
This is due to the factor $(1-r)^2$, which shows up in the expansion of Eq.~(\ref{eq:qqbaroctetta}). 
However, starting from $\ord(\beta^5)$ (or $\ord(\beta^3)$ 
in the color-summed result of Eq.~(\ref{eq:fqq00})), 
there is no such factor and one would create an artificial divergence, if 
one inserts the full LO scaling function into Eq.~(\ref{eq:fqq10}). 
On the other hand, it is clear that in the exact NLO result, the function 
$A^{\qqbar}_{{\bf 8}_a}(r)$ should also possess terms depending on $\beta$. 
These cancel the problematic higher order terms when multiplied with the
Born function. Moreover, they give additional contributions to the NLO cross 
section, which are not treated by the ansatz (\ref{eq:hardfunc}).
Note, that in the gluon-fusion channel, Eq.~(\ref{eq:C1ggI}) is free of
artificial divergences and the function $A_{\bf I}^{gg}(r)$ is well defined for all $r>0$.

Within the regularization scheme $\overline{\rm DR}$ underlying 
Eqs.~(\ref{eq:C1ggI}) and~(\ref{eq:C1qq8}), the strong coupling constant is 
understood to be evaluated at a hard scale where all squark flavors as 
well as the gluino contribute within the virtual corrections. 
Conventional QCD computations on the other hand employ the $\overline{\rm MS}$-scheme, 
and so does the program \texttt{Prospino}~\cite{Beenakker:1996ed}. 
In order to compare to the numerical output of \texttt{Prospino} 
in the $\overline{\rm MS}$-scheme, 
we have to perform a scheme transformation and decouple the SUSY particles as well 
as the top-quark from the spectrum. 
The necessary change of the renormalization scheme for $\as$ and the decoupling 
can easily be done with the help of formulae given in Ref.~\cite{Harlander:2005wm}. 
Assuming that the top-quark is lighter than all sparticles, 
we have to add the following shift to the NLO scaling functions in Eq.~(\ref{eq:definef})
\begin{eqnarray}
\label{eq:scheme}
\Delta 
\left.\left(
f_{ij,\,{\bf I}}^{(10)}
+
f_{ij,\,{\bf I}}^{(11)} L_\mu
\right)
\right|_{\rm \overline{DR}(MSSM)\to \overline{MS}(n_l=5)}
&=& 
\frac{f^{(00)}_{ij,\,{\bf I}}}{4\pi^2}\* \left(
\frac{1}{2}  - \frac{1}{3}\*L_{t\widetilde g} - \frac {n_f}{6}\*\ln(r)
+ \frac{4}{3}\*L_\mu + \frac{n_f}{6}\*L_\mu
\right),
\end{eqnarray}
where $n_l$ denotes the number of light (massless) quark flavors. 
This leads to the one-loop matching constants 
in the $\overline{\rm MS}$-scheme with a total of $n_f$ quark flavors 
($n_f=n_l+1$), 
\begin{eqnarray}
\label{eq:C1ggI-MS}
C_{1,\,\bf I}^{gg\,\overline{\rm MS}} &=& 
C_{\bf I}\*\left(4 + \ln(2) -\frac{\pi^2}{8}\right) 
        + 36 - 6\,\*\ln^2(2) -\pi^2
        + \frac{n_f}{18}\,\*A_{\bf I}^{gg}(r)
\, ,
\\
\label{eq:C1qq8-MS}
C_{1,\,{\bf 8}_a}^{\qqbar\,\overline{\rm MS}} &=&
n_f\*\left(\ln(2)- \frac{5}{9}\right) + \frac{181}{6}
          - \frac{8}{3}\,\*\ln^2(2) - \frac{43}{36}\,\*\pi^2 
          - \frac{1}{3}\,\*L_{t\widetilde g} - \frac{n_f}{6}\,\*\ln(r)
          + A^{\qqbar}_{{\bf 8}_a}(r)          
\, .
\end{eqnarray}

We remark here, that the gluino bound state computation of Ref.~\cite{Kauth:2011vg} 
has been performed in the limit $m_t\to 0$ wherever possible. 
Thus, the dependence on $m_t$ in Eq.~(\ref{eq:C1ggI}) for $C_{1,\,{\bf I}}^{gg\,\overline{\rm DR}}$
in the $\overline{\rm DR}$-scheme and in Eq.~(\ref{eq:C1qq8-MS}) for 
$C_{1,\,{\bf 8}_a}^{\qqbar\,\overline{\rm MS}}$ in the $\overline{\rm MS}$-scheme is only logarithmic.
The NLO QCD corrections to the inclusive cross section~\cite{Beenakker:1996ch} 
coded in the program \texttt{Prospino}~\cite{Beenakker:1996ed}, 
on the other hand, account for the complete dependence on $m_t$. 

In order to cross check our analytic results in Eqs.~(\ref{eq:C1ggI-MS}) and~(\ref{eq:C1qq8-MS}), 
we numerically extract the one-loop hard matching coefficients of the color-summed NLO scaling functions 
at their threshold from \texttt{Prospino}, cf. Eq.~(\ref{eq:Cgg1}).
For our numerical analysis, we set the squark masses to $600\,$GeV and vary
the gluino mass between $100\,\GeV$ and $2\,\TeV$ in steps of $100\,$GeV.
The top-quark mass is set to $175\,$GeV in the on-shell scheme.
According to Eq.~(\ref{eq:fgg00}), the LO scaling function of the gluon-fusion 
channel does not depend on any of the SUSY masses and 
the top-quark mass dependence of $C_{1,\,{\bf I}}^{gg\,\overline{\rm MS}}$ has
canceled in Eq.~(\ref{eq:C1ggI-MS}).
Thus, for the case of gluon-fusion, we find agreement with our analytic result 
within a few per mill over the whole range of input values.
For $\qqbar$-annihilation on the other hand, we encounter a dependence on the 
mass ratio $r$ in Eq.~(\ref{eq:def-variables}) for equal squark masses and 
we expect deviations due to finite contributions proportional to the top-quark mass. 
We find differences between the expression for 
$C_{1,\,{\bf 8}_a}^{\qqbar\,\overline{\rm MS}}$ based on Ref.~\cite{Kauth:2011vg} 
in Eq.~(\ref{eq:C1qq8-MS}) and the result extracted from \texttt{Prospino}, 
which amount to the order of a few per cent especially for mass ratios $r>1$.
Altogether, this constitutes an important cross check, both of our derivation 
and of the original computation of the NLO corrections in Ref.~\cite{Beenakker:1996ed}.
Moreover, as already noted, the gluon channel is dominant for collider physics
predictions at the LHC. Therefore, we are able to provide extremely accurate 
predictions for the gluino pair production cross section in the threshold region.

\bigskip

We are now in the position to present the NNLO cross section in the threshold limit exact to NNLL accuracy. 
All coefficients of the threshold logarithms $\ln^n(\beta)$ at NNLO 
can be calculated from the resummation formula (\ref{eq:sigmaNres}) 
with the exponent (\ref{eq:GNexp}) after an inverse Mellin transformation 
and with the knowledge of the one-loop matching coefficients 
$C_{1,\,{\bf I}}^{ij}$ in Eqs.~(\ref{eq:C1ggI-MS}) and~(\ref{eq:C1qq8-MS}).
Note, that at $\ord(\as^2)$ we only keep logarithmically enhanced 
terms proportional to powers of $\ln(\beta)$ as well as Coulomb corrections in the following.
The two-loop matching coefficients $C_{2,\,{\bf I}}^{ij}$ defined in analogy 
to Eqs.~(\ref{eq:fgg10}) and~(\ref{eq:fqq10}) are presently unknown and we set
them to zero in the results for the NNLO cross section in the threshold limit below.
The determination of the two-loop hard constants $C_{2,\,{\bf I}}^{ij}$ 
would require a complete NNLO calculation, which is beyond the scope of the
present study.

For the gluon-fusion channel we obtain in this way 
the NNLO scaling functions in the threshold approximation as,
\begin{eqnarray}
\label{eq:fgg20}
f^{(20)}_{gg,\,{\bf I}} &=& \frac{f^{(00)}_{gg,\,{\bf I}}}{(16\*\pi^2)^2}\*
\Biggl[
\frac{4\*D_{\bf I}^2\*\pi^4}{3\*\beta^2} 
+ \frac{D_{\bf I}\*\pi^2}{\beta} \*\biggl\{ 
   -  192\*\ln^2(\beta) + \left(44 + 16\*C_{\bf I} 
   - \frac{8}{3}\*n_l - 192\*\ln(2) \right)\* \ln(\beta)
\\
&& - 8\*C_{1,\,{\bf I}}^{gg} + \frac{1090}{3} + 16\*C_{\bf I} 
 + \frac{4}{3}\*n_l\*\left(\frac{5}{3} - 2\*\ln(2)\right)
 + 44\*\ln(2) + 8 \*C_{\bf I}\*\ln(2) - 48\*\ln^2(2)
 - 16\*\pi^2 \biggr\} 
\nonumber\\
&& 
 + 4608\*\ln^4(\beta) 
 + \biggl\{ -19840 - 768\*C_{\bf I} + \frac{256}{3}\*n_l + 27648\*\ln(2) \biggr\}\*\ln^3(\beta)  
\nonumber\\
&& 
 + \biggl\{
  384\,\*C_{1,\,{\bf I}}^{gg} + 43232 + 1712\*C_{\bf I} + 32\*C_{\bf I}^2 
 + n_l\*\Big{(}-\frac{1088}{3} - \frac{32\*C_{\bf I}}{3} 
 + 384\*\ln(2)\Big{)}
\nonumber\\
&& - 89280\*\ln(2) - 3456\*C_{\bf I}\*\ln(2) + 62208\*\ln^2(2) 
  - 2400\*\pi^2\biggr\}\*\ln^2(\beta) 
\nonumber\\
&& + \biggl\{-\frac{262624}{3}- \frac{6584}{3}\*C_{\bf I} + 
    \Big{(}-768 - 32\*C_{\bf I} + 1152\*\ln(2)\Big{)}\*C_{1,\,{\bf I}}^{gg} 
\nonumber\\
&& + n_l\*\left(\frac{6976}{9} + \frac{368}{9}\*C_{\bf I}
    - 1088\*\ln(2) - 32\*C_{\bf I}\*\ln(2) + 576\*\ln^2(2) - 
       32\*\pi^2\right)
\nonumber\\
&& + 129696\*\ln(2) + 5136\*C_{\bf I}\*\ln(2) + 
    96\*C_{\bf I}^2\*\ln(2) - 133920\*\ln^2(2) - 5184\*C_{\bf I}\*\ln^2(2) 
\nonumber\\
&& + 62208\*\ln^3(2) + 5328\*\pi^2 + 200\*C_{\bf I}\*\pi^2 - 7200\*\ln(2)\*\pi^2
  + 33264\*\zeta_3 - 48\*C_{\bf I}\*\zeta_3
\nonumber\\
&& + 16\*\pi^2\* D_{\bf I} \* \Bigl(3-2\* D_{\bf I}\* (1 + v_{\rm spin})\Bigr)
\biggr\}\*\ln(\beta)
\nonumber
+ C_{2,\,{\bf I}}^{gg} 
\Biggl]\, ,
\end{eqnarray}
with $D_{\bf I}$ and $C_{\bf I}$ given in Eqs.~(\ref{eq:CI}), (\ref{eq:DI}) and 
$C_{1,\,\bf I}^{gg}$ in Eq.~(\ref{eq:C1ggI-MS}) for $\overline{\rm MS}(n_l=5)$.
$\zeta_i$ denote the values of the Riemann zeta function. 
Likewise, for quark-antiquark annihilation, we find
\begin{eqnarray}
\label{eq:fqq20}
f^{(20)}_{\qqbar,\,{\bf 8}_a} &=& \frac{f^{(00)}_{\qqbar}}{(16\*\pi^2)^2}\*
\Biggl[
\frac{3\*\pi^4}{\beta^2} 
+ \frac{\pi^2}{\beta}\*\biggl\{
  128\*\ln^2(\beta) 
+ \left(-138 + 4\*n_l + 128\*\ln(2)\right)\*\ln(\beta) 
+ 12\,\*C_{1,\,{\bf 8}_a}^{\qqbar} 
\\
&& - 297 + n_l\*\Big{(}-\frac{10}{3} + 4\*\ln(2)\Big{)} - 102\*\ln(2) + 32\*\ln^2(2) 
  + \frac{32\*\pi^2}{3}
  \biggr\} 
\nonumber\\
&& 
+ \frac{8192}{9}\*\ln^4(\beta) 
+ \frac{512}{27}\*\biggl\{-279 + 288\*\ln(2) + 2\*n_l\biggr\}\*\ln^3(\beta) 
+ \biggl\{12976 + \frac{512}{3}\,\*C_{1,\,{\bf 8}_a}^{\qqbar}
\nonumber\\
&& 
+ n_l\*\Big{(}-\frac{5216}{27} + \frac{512}{3}\*\ln(2)\Big{)} 
- 23808\*\ln(2) + 12288\*\ln^2(2) - \frac{4480}{9}\*\pi^2\biggr\}\* \ln^2(\beta) 
\nonumber\\
&& 
+ \biggl\{\Bigl(-\frac{1312}{3}
+ 512\*\ln(2)\Bigr)\*C_{1,\,{\bf 8}_a}^{\qqbar} -\frac{667624}{27}
+ n_l\*\Bigl(\,\frac{37840}{81} - \frac{5216}{9}\*\ln(2) 
\nonumber\\ 
&& 
+ 256\*\ln^2(2) - \frac{128}{9}\*\pi^2\Bigr) + 38928\*\ln(2) 
- 35712\*\ln^2(2) + 12288\*\ln^3(2) 
\nonumber\\
&& 
+ \frac{13592}{9}\*\pi^2 - \frac{4480}{3}\*\ln(2)\*\pi^2 
+ \frac{60080}{9}\*\zeta_3 
+ 16\*\pi^2\* D_{\bf I} \* \Bigl(3-2\* D_{\bf I}\* (1 + v_{\rm spin})\Bigr)
\biggr\}\*\ln(\beta)
\nonumber
+ C_{2,\,{\bf 8}_a}^{\qqbar}
\Biggr]
\, ,
\end{eqnarray}
with the $\overline{\rm MS}$-scheme result for $C_{1,\,{\bf 8}_a}^{\qqbar\,\overline{\rm MS}}$ 
from Eq.~(\ref{eq:C1qq8-MS}).
The results in Eqs.~(\ref{eq:fgg20}) and (\ref{eq:fqq20}) agree with Ref.~\cite{Beneke:2009ye} 
where the approximate NNLO cross section at threshold has been computed for massive colored particle 
production in an arbitrary $SU(3)_{\mathrm{color}}$ representation of the final state.
In particular, they also contain subleading NNLO Coulomb terms and the non-relativistic kinetic-energy 
corrections, which do not follow directly from the 
resummed cross section~(\ref{eq:sigmaNres}), but have to be determined 
from matching to explicit NNLO 
computations~\cite{Czarnecki:1997vz,Beneke:1999qg,Czarnecki:2001gi,Pineda:2006ri}.
The latter ones are given by terms proportional to $D_{\bf I} \* (3-2\* D_{\bf I}\* (1 + v_{\rm spin}))\ln(\beta)$ 
in Eqs.~(\ref{eq:fgg20}) and (\ref{eq:fqq20}) and depend on 
spin configuration of the $\gluinopair$ final state through the quantity $v_{\rm spin}\,$. 
For the gluino pair in a spin-singlet configuration 
as realized for the $S$-wave in the $\lbrace {\bf 1,8,27} \rbrace$ symmetric
color representations of the gluon-fusion channel it takes the value $v_{\rm spin}=0$. 
For a spin-triplet as in the antisymmetric octet representation of the 
$q\bar{q}$ channel we have $v_{\rm spin}=-2/3$.

For direct comparison, we also present here the 
one-loop matching coefficients $C_{1,\,{\bf I}}^{gg}$ and $C_{1,\,{\bf 8}_a}^{\qqbar}$ 
in the notation of Ref.~\cite{Beneke:2009ye} (cf. $C_X^{(1)}$ in Eq.~(A.2) of that reference).
Using Eqs.~(\ref{eq:fgg10}), (\ref{eq:fqq10}), (\ref{eq:C1ggI-MS}) and (\ref{eq:C1qq8-MS}), 
we find for $\mu_f=\mu_r=\mu$ 
\begin{eqnarray}
\label{eq:ReC}
\text{Re}[C_{gg\to\gluinopair}^{(1)}] &=& 
2\,\*C_{1,\,{\bf I}}^{gg} - 6\*\,C_{\bf I}
- C_A\*\left(32 - \frac{11}{6}\*\,\pi^2 - 4\*\,\ln(2)\* L_\mu +  L_\mu^2\right)
- C_{\bf I}\*\,L_\mu
\, ,
\\
\text{Re}[C_{\qqbar\to\gluinopair}^{(1)}] &=& 
2\,\*C_{1,\,{\bf 8}_a}^{\qqbar} - 6\*\,C_A
- C_F\*\left(32 - \frac{11}{6}\*\,\pi^2 - 4\*\,\ln(2)\*L_\mu +  L_\mu^2\right)
+ \left(\frac{14}{3} - \frac{2}{3}\*\,n_f\right)\*L_\mu
\, ,\qquad
\end{eqnarray}
which displays an additional dependence on the renormalization scale due to the
particular normalization of Ref.~\cite{Beneke:2009ye}.
Also note, that our choice $n_f=6$ corresponds to $n_l+1$ in the notation of 
Ref.~\cite{Beneke:2009ye}.

\bigskip

For completeness we briefly list all functions governing the scale dependence
up to NNLO in the gluino pair production cross section.
These can be computed by standard renormalization group methods 
(see e.g.,~\cite{Langenfeld:2009wd}) 
in terms of coefficients $\beta_l$ of the QCD beta-function and 
the splitting functions $P_{ij}$ which govern the PDF evolution.
For the hard functions $f_{ij}^{(11)}$ in the $\overline{\rm MS}$-scheme 
we have 
\begin{eqnarray}
\label{eq:f11gen}
f_{ij,\,{\bf I}}^{(11)} = \frac{1}{16\pi^2}\*\left(
  2\* \beta_0 \*f_{ij,\,{\bf I}}^{(00)} - f_{kj,\,{\bf I}}^{(00)}\otimes P_{ki}^{(0)} 
- f_{ik,\,{\bf I}}^{(00)}\otimes P_{kj}^{(0)}\right)\,,
\end{eqnarray}
where $\otimes$ denotes the standard Mellin convolution, 
and repeated indices imply summation over admissible partons.
The coefficients $\beta_l$ and $P_{ij}^{(k)}$ are taken in an expansion in 
powers of $\as/(4\pi)$ as in the normalization~(\ref{eq:fctexpand}) 
(see Refs.~\cite{Moch:2004pa,Vogt:2004mw}).
The explicit expressions near threshold read 
\begin{eqnarray}
\label{eq:f11gg}
f^{(11)}_{gg,\,{\bf I}} &=& 
-\frac{f^{(00)}_{gg,\,{\bf I}}}{16\pi^2}\,
\Big{(}24\*\ln(8\*\beta^2) - 48 - 24\*\ln(2)\Big{)}
\, ,\\[3mm]
\label{eq:f11qq}
f^{(11)}_{\qqbar,\,{\bf 8}_a} &=& 
-\frac{f^{(00)}_{\qqbar,\,{\bf 8}_a}}{16\pi^2}\,
\left(\frac{32}{3}\*\ln(8\*\beta^2) -\frac{106}{3} -\frac{32}{3}\*\ln(2)+\frac{4}{3}\*n_l\right)
\, .
\end{eqnarray}
Recall, that $n_l$ denotes the number of light quark flavors. 
Likewise, at NNLO, the scale dependent part can be calculated by evaluating 
\begin{eqnarray}
\label{eq:f21}
f_{ij}^{(21)} &=&
\frac{1}{(16\pi^2)^2}\*\left(
  2\* \beta_1\* f_{ij}^{(00)} - f_{kj}^{(00)}\otimes P_{ki}^{(1)}
  - f_{ik}^{(00)}\otimes P_{kj}^{(1)}\right)
\\
&& 
+ \frac{1}{16\pi^2}\*\left(
  3 \*\beta_0 \*f_{ij}^{(10)}
  -f_{kj}^{(10)}\otimes P_{ki}^{(0)}
  - f_{ik}^{(10)}\otimes
  P_{kj}^{(0)}\right)
\, ,
\nonumber\\[2mm]
\label{eq:f22}
f_{ij}^{(22)} &=&
\frac{1}{(16\pi^2)^2}\*\left(
  f_{kl}^{(00)}\otimes P_{ki}^{(0)}\otimes P_{lj}^{(0)}
  +\frac{1}{2} f_{in}^{(00)}\otimes P_{nl}^{(0)}\otimes P_{lj}^{(0)}
  +\frac{1}{2} f_{nj}^{(00)}\otimes P_{nk}^{(0)}\otimes P_{ki}^{(0)}
	\right.
\\[2mm]
&& \hspace{18mm}
\left. + 3 \*\beta_0^2  \*f_{ij}^{(00)}
  - \frac{5}{2}\*\beta_0 \*f_{ik}^{(00)}\otimes P_{kj}^{(0)}
  - \frac{5}{2}\*\beta_0 \*f_{kj}^{(00)}\otimes P_{ki}^{(0)}
\right)
\, ,
\nonumber
\end{eqnarray}
where we have suppressed the color indices ${\bf I}$. 
Inserting the threshold approximations of the splitting functions
and the NLO scaling functions,
we obtain the desired results for $ij=gg, \qqbar$ 
near threshold, which explicitly read 
\begin{eqnarray}
\label{eq:fgg21}
f^{(21)}_{gg,\,{\bf I}} &=& 
\frac{f^{(00)}_{gg,\,{\bf I}}}{(16\pi^2)^2}\*\Biggl[
-4608\*\ln^3\left(\beta\right) +
 \biggl(14880 + 384\*C_{\bf I} - 18432\*\ln(2) - 64\*n_l\biggr)
 \*\ln^2\left(\beta\right)
\\
&& + \biggl(\frac{96\*\pi^2}{\beta}\*\,D_{\bf I} - 21616  
  - 472\*C_{\bf I} + 960\* C_{\bf I}\*\ln(2) - 192\* C_{1,\,{\bf I}}^{gg}
\nonumber\\
&& + n_l\*\,\frac{4}{3}\*\Big{(} 136 + 4\*C_{\bf I} - 144\*\ln(2)\Big{)} + 
    40032\*\ln(2) - 24192\*\ln^2(2) + 1200\*\pi^2\biggr)\*\ln(\beta)
\nonumber\\
&& - \frac{22\*\pi^2}{\beta}\*\,D_{\bf I} + 19516 
 + 12\*C_{\bf I}\*\left(32 - 59\* \ln(2) + 48\*\ln^2(2) - 4\*\pi^2\right) 
 + 236\*C_{1,\,{\bf I}}^{gg}
\nonumber\\
&&  + n_l\*\,\frac{4}{3}\*\left( \frac{\pi^2}{\beta}\*\,D_{\bf I} 
     - 43 + 6\*C_{\bf I}\*\ln(2) - 2\*C_{1,\,{\bf I}}^{gg} + 184\*\ln(2) - 
    108\*\ln^2(2)\right) - 192\*C_{1,\,{\bf I}}^{gg}\*\ln(2) 
\nonumber\\
&& - 31888\*\ln(2) + 26568\*\ln^2(2) - 10368\*\ln^3(2) + 
 1776\*\ln(2)\*\pi^2 
 - 1200\*\pi^2 - 8280\*\zeta_3 
\Biggr]\,,
\nonumber\\[2ex]
\label{eq:fgg22}
f^{(22)}_{gg,\,{\bf I}} &=& 
\frac{f^{(00)}_{gg,\,{\bf I}}}{(16\pi^2)^2}\*\Biggl[
1152\*\ln^2\left(\beta\right) 
+ \biggl(- 2568 + 16\,\* n_l + 2304\*\ln (2)\biggr)\*\ln(\beta)\\
&& + 16\,\*n_l\*\Big(-1 + \ln(2)\Big)
+ 2568 - 2568\*\ln(2) + 1152\*\ln^2(2) -144\*\pi^2 
\nonumber
\Biggr]\,,
\nonumber\\[2ex]
\label{eq:fqq21}
f^{(21)}_{\qqbar,8a} &=&
\frac{f^{(00)}_{\qqbar}}{(16\*\pi^2)^2}\*\Biggl[
-\frac{8192}{9}\*\ln^3\left(\beta\right) + 
 \frac{128}{9}\*\biggl(303 - 256\*\ln(2) - 6\*\,n_l \biggr)\*
  \ln^2\left(\beta\right)
 \\
&& - \biggl(\frac{64\*\pi^2}{\beta} + 
    \frac{8}{27}\*\Big{(}288\*C_{1}^{\qqbar} + 24849 + n_l\*\Big{(}-818 
   + 864\*\ln(2)\Big{)} - 39696\*\ln(2)
 \nonumber\\
&& + 16128\*\ln^2(2) - 840\*\pi^2\Big{)}\biggr)\*\ln(\beta) 
+ \frac{75\*\pi^2}{\beta} + 
 \frac{4}{3}\*C_{1}^{\qqbar}\*\left(139 - 64\*\ln(2)\right) 
+ \frac{14416}{3}
 \nonumber\\
&& - n_l\*\left(\frac{6\*\pi^2}{\beta} + 8\*C_{1}^{\qqbar} + 
    \frac{4}{27}\*\left(325 - 2374\*\ln(2) + 1296\*\ln^2(2) 
   - 8\*\pi^2\right)\right)
 \nonumber\\
&& - \frac{4}{9}\*\Big{(}24313\*\ln(2) - 17880\*\ln^2(2) + 4608\*\ln^3(2) 
 - 816\*\ln(2)\*\pi^2 + 732\*\pi^2
 + 3560\*\zeta_3\Big{)}
\Biggr]\,,
\nonumber\\[2ex]
\label{eq:fqq22}
f^{(22)}_{\qqbar,8a} &=& 
\frac{f^{(00)}_{\qqbar}}{(16\*\pi^2)^2}\*\Biggl[
\frac{2048}{9}\*\ln^2(\beta)
+ \frac{4}{9}\*\biggl(80\,\*n_l - 1960 + 1024\*\ln(2)\biggr)\*\ln(\beta)
+\frac{4}{3}\*n_l^2
 \\
&& + n_l\*\,\frac{4}{9}\*\Big(-149 + 80\*\ln(2)\Big)
+ \frac{1}{9}\*\left(9415 - 7840 \*\ln(2) + 
   2048\*\ln^2(2) - 256\*\pi^2\right)
\nonumber
\Biggr]\,. 
\end{eqnarray}

\bigskip

\begin{figure}[h!]
  \centering
    \scalebox{0.7}{\includegraphics{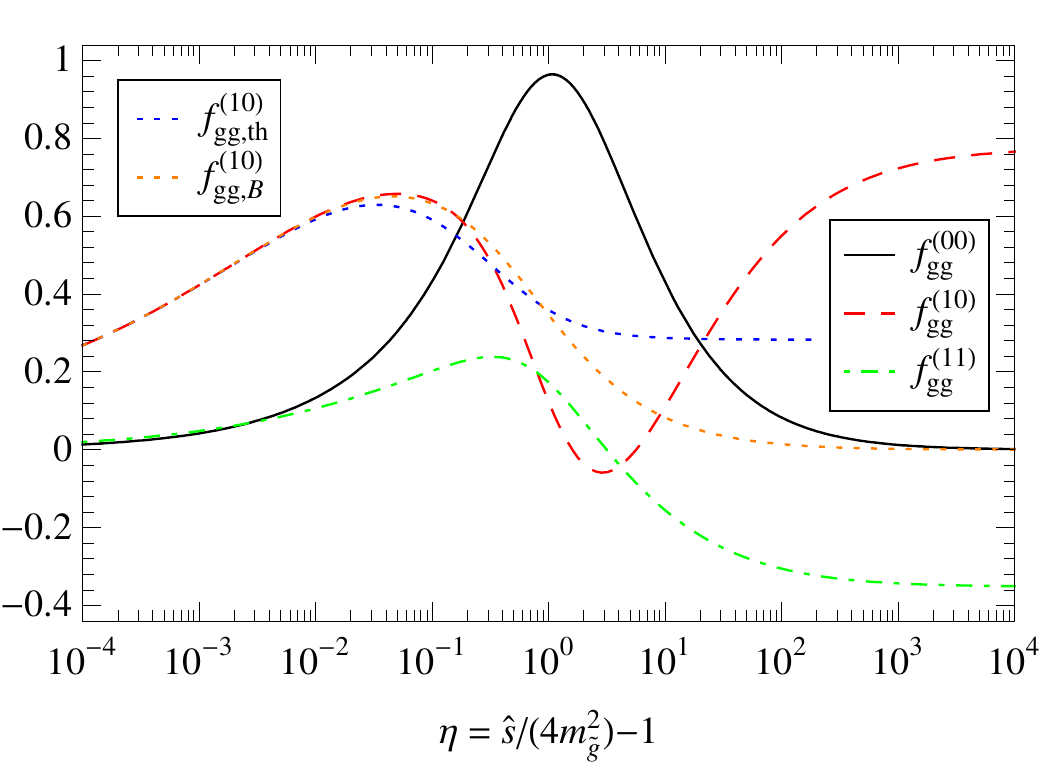}}
    \hspace*{5mm}
    \scalebox{0.7}{\includegraphics{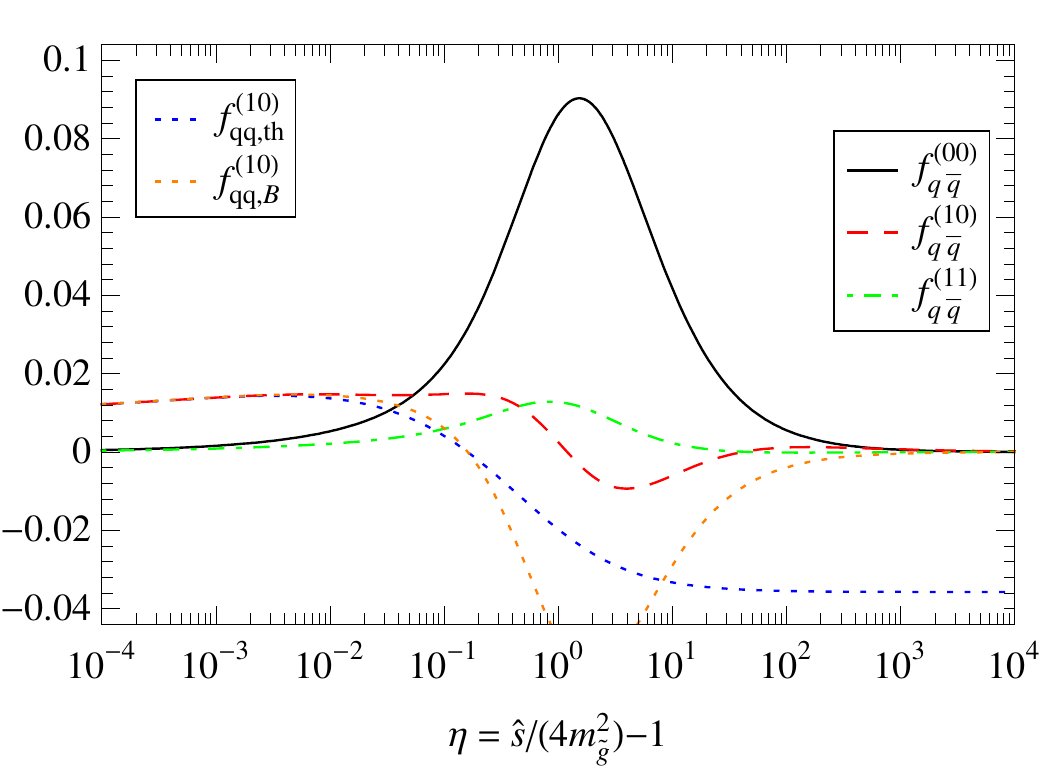}}
  \vspace*{8mm}

  \scalebox{0.7}{\includegraphics{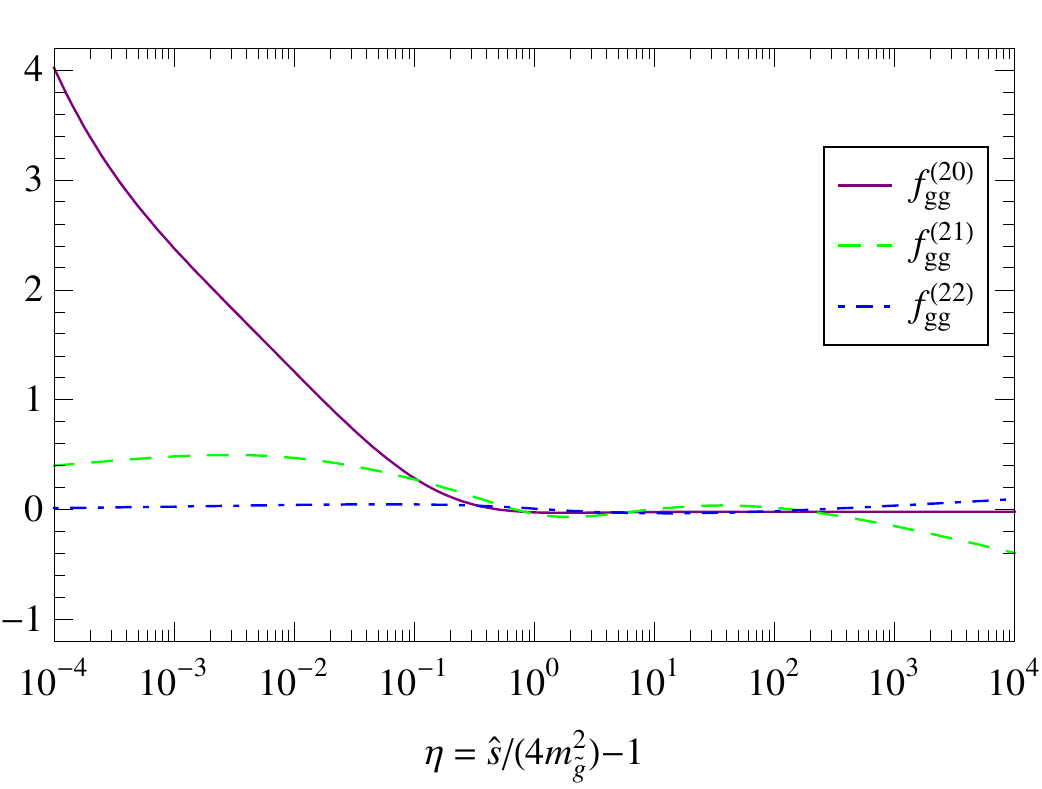}}
  \hspace*{5mm}
  \scalebox{0.7}{\includegraphics{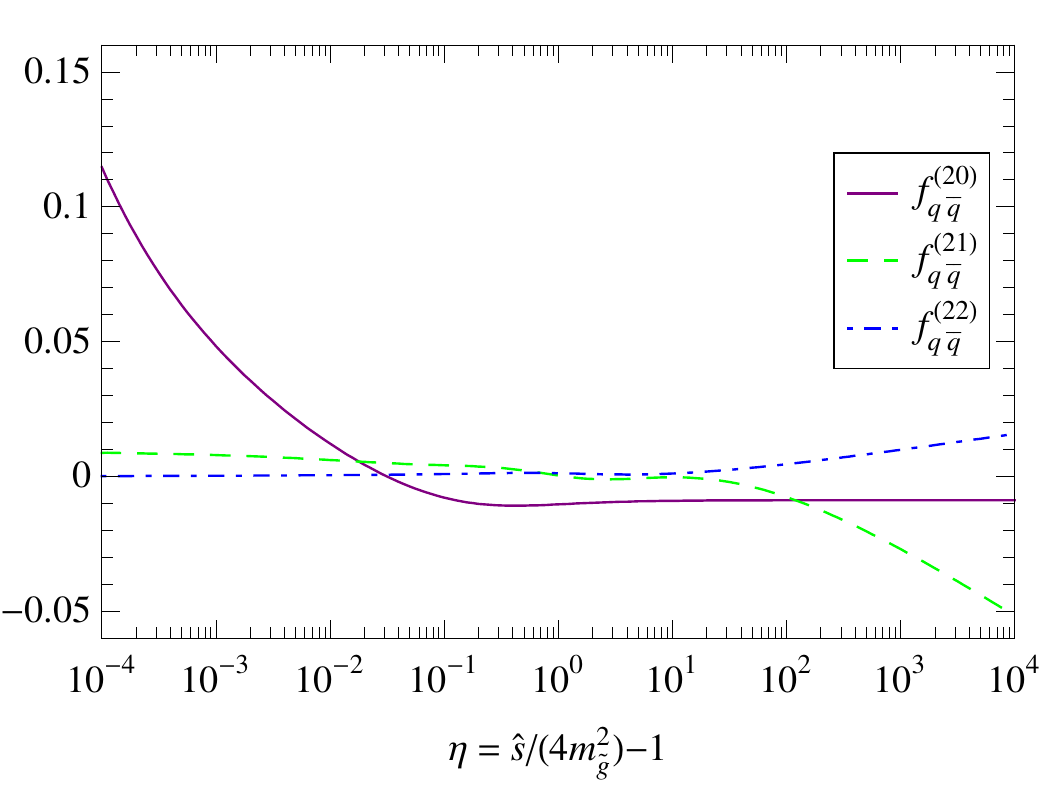}}
 \vspace*{8mm}
  
  \scalebox{0.7}{\includegraphics{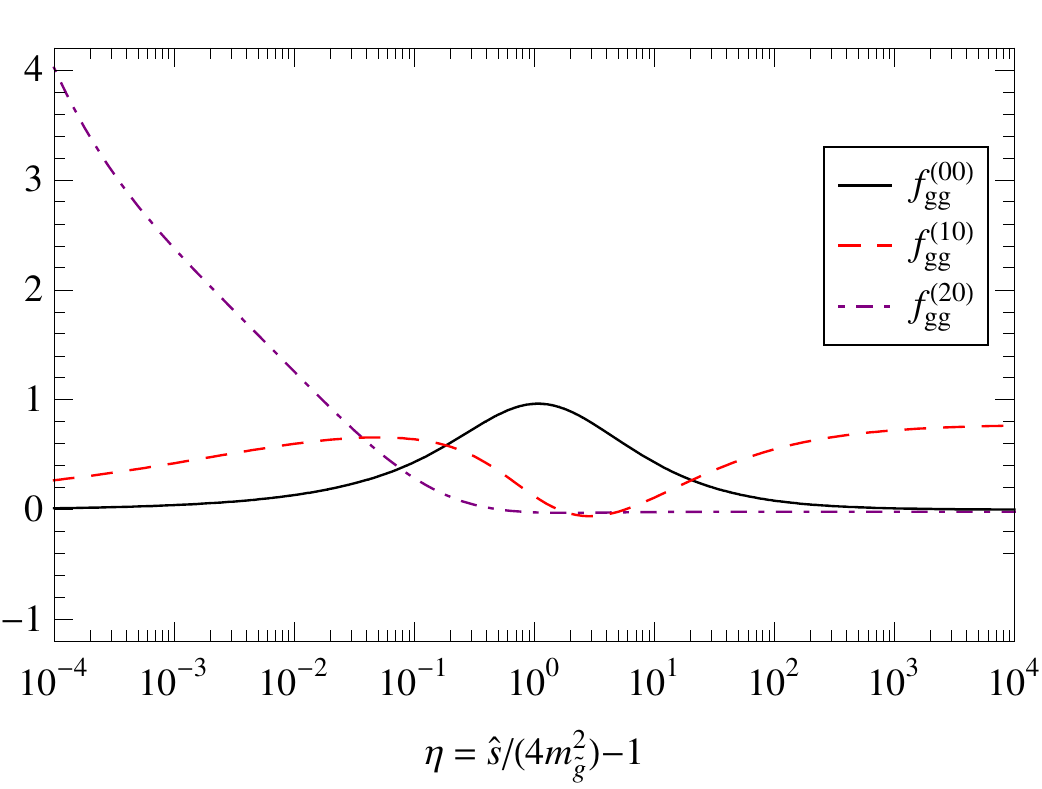}}
  \hspace*{5mm}
  \scalebox{0.7}{\includegraphics{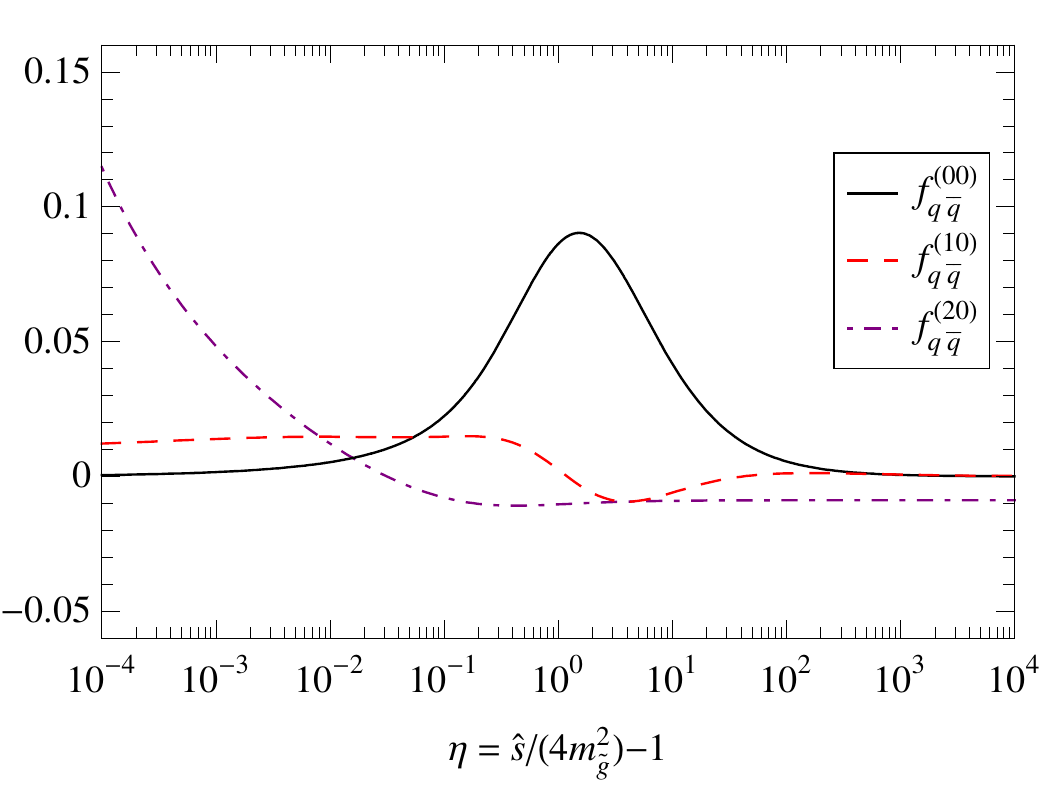}}
 \vspace*{5mm}
 \caption{\small{The scaling functions $f^{(ij)}_{gg}$ and $f^{(ij)}_{\qqbar}$ 
     with $i = 0,1,2$ and $j\le i$ in the $\overline{\rm MS}$-scheme. 
     The masses are $\mg = 750\,\GeV$ and $\mq = 600\,\GeV$.}}
  \label{fig:scalingfunctions}
\end{figure}

In Fig.~\ref{fig:scalingfunctions}, we plot the color-summed
NLO and NNLO scaling functions. 
For comparison we also show the exact LO results given in Eqs.~(\ref{eq:fgg00}) and (\ref{eq:fqq00}). 
We use a gluino mass $\mg = 750\GeV$ and squark masses $\mq = 600\GeV$ which correspond to $r = 0.64$.
In the gluon-fusion channel, the dependence on $r$ starts at NLO and is rather weak in the threshold region. For 
$\qqbar$-annihilation however, one has a stronger dependence 
already starting at LO. 
The NLO results $f^{(10)}_{ij}$ and $f^{(11)}_{ij}$ in Fig.~\ref{fig:scalingfunctions} are exact.
Similarly, as the results for NNLO scale dependent functions $f^{(21)}_{ij}$ and $f^{(22)}_{ij}$ 
are based on Eqs.~(\ref{eq:f21}) and (\ref{eq:f22}), 
they are also exact at all energies even away from threshold.
For the genuine NNLO contributions $f^{(20)}_{ij}$ we plot 
our new results~(\ref{eq:fgg20}) and (\ref{eq:fqq20}). 
The threshold approximation for the latter functions could, in principle, be improved 
by adding constraints imposed by the high-energy factorization, 
see~\cite{Moch:2012mk} for related studies in top-quark hadro-production.  
However, given the large gluino masses currently considered, this is not
immediately relevant for phenomenology at current and foreseeable LHC energies.

The range of validity of the threshold expansion is demonstrated for the NLO scaling functions in the upper two figures. Here, we plot in addition the approximated results which contain only threshold enhanced terms and constants (subscript th), and the improved threshold approximations, where the exact Born terms are inserted into eqs. (\ref{eq:fgg10}) and (\ref{eq:fqq10}) (subscript B).
In the latter case, the curves follow the behavior of the Born terms at high velocities, which tend to zero for $\beta\to 1$. In the former case, an offset arises which, for $\qqbar$-annihilation, depends on $r$.
In the gluon-fusion channel, the formulae work very well up to $\eta\approx 0.4$,
which corresponds to $\beta=\sqrt{\eta/(1+\eta)}\approx 0.53$.
For quark-antiquark annihilation, high accuracy is guaranteed up to $\eta\approx 10^{-2}$ ($\beta\approx 0.1$). 

\begin{figure}
 \centering
\scalebox{0.7}{\includegraphics{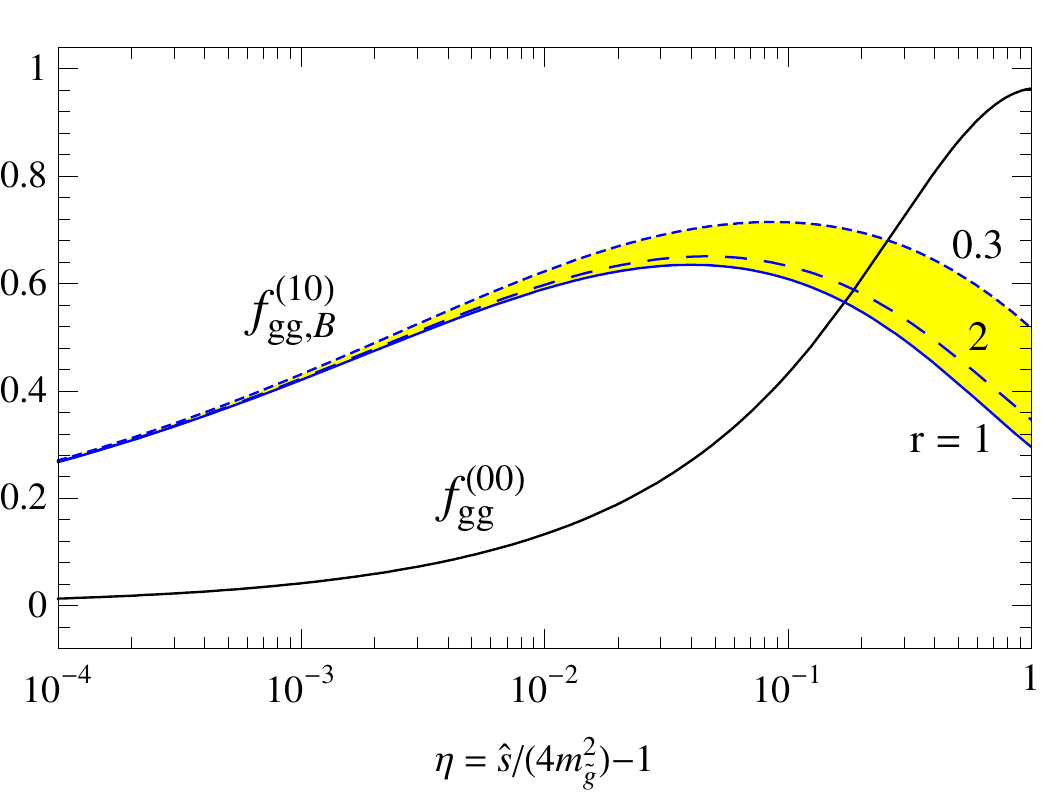}}
\caption{
\small{ 
  Scaling function $f^{(10)}_{gg,B}$ (threshold approximation with exact Born function)
  for $0.3<r<2$.}
}
\label{fig:fggvar}
\end{figure}

It should be stressed further that the scaling functions in the $gg$~channel exceed those 
of the $\qqbar$~channel by about one order of magnitude as shown in Fig.~\ref{fig:scalingfunctions}. 
Keeping in mind that also the parton luminosity at a proton-proton collider
such as the LHC favors the channel with initial state gluons over the one with quarks in the TeV-regime, we conclude that gluon-fusion is by far the dominant source for $\gluinopair$-production at the LHC. 
Thus, the theory predictions of the inclusive $\gluinopair$ hadro-production cross section are mainly governed by the gluino mass and are rather insensitive to the squark masses. For illustration, we also plot the NLO scaling function (\ref{eq:fgg10}) for different values of $r$ in Fig.~\ref{fig:fggvar}. Its weak dependence on $r$ is minimized for equal squark and gluino masses ($r=1$).
Recall that the LO cross section in the gluon-fusion channel does not 
depend on $r$.

\section{Hadronic cross section}
\label{sec:hadronic}

Here we discuss the total hadronic cross section, which is obtained by convoluting the 
partonic scaling functions with the PDFs, see Eq.~(\ref{eq:totalcrs}).
For the numerical results we keep the threshold enhanced channels $gg$ and $\qqbar$ 
at all orders up to NNLO, while we consider only the NLO contributions for $gq$, 
which are the leading contributions of this channel.
As already discussed in Sec.~\ref{sec:partonic}, 
at the hadronic level, the $gg$-channel accounts for the largest part, 
whereas the contribution of the $\qqbar$-channel is a few percent of the $gg$-channel, only.

For reasons of convenience, the computation of the hadronic cross sections
employs a grid in the $\mg$ - $\mq$ -plane for the scale independent scaling functions
$f^{(10)}_{ij}$, $ij = gg, \qqbar, gq$, which has been extracted from \texttt{Prospino}.
This grid has already been applied in the numerical check of the one-loop matching constant 
$\Cgg$ and $\Cqq$ (see the previous Sec.~\ref{sec:partonic}).
For the hadronic cross section computation considered here, 
these scaling functions are used to calculate the 
exact scale dependent scaling functions $f^{(11)}_{ij}$, $f^{(21)}_{ij}$ and $f^{(22)}_{ij}$ 
from the renormalization group Eqs.~(\ref{eq:f11gen}), (\ref{eq:f21}) and (\ref{eq:f22}).
Using these results together with the threshold approximation for
$f^{(20)}_{ij}$ from Eqs.~(\ref{eq:fgg20}) and (\ref{eq:fqq20}) defines 
the theory predictions at approximate NNLO (dubbed NNLO$_{\text{approx}}$
in the sequel) to be used in our phenomenological studies.
Further improvements based on the evaluation of the resummation formula~(\ref{eq:sigmaNres}) to account for threshold logarithms at all orders to NNLL accuracy are postponed to future work.

\begin{figure}
 \centering
\scalebox{0.35}{\includegraphics{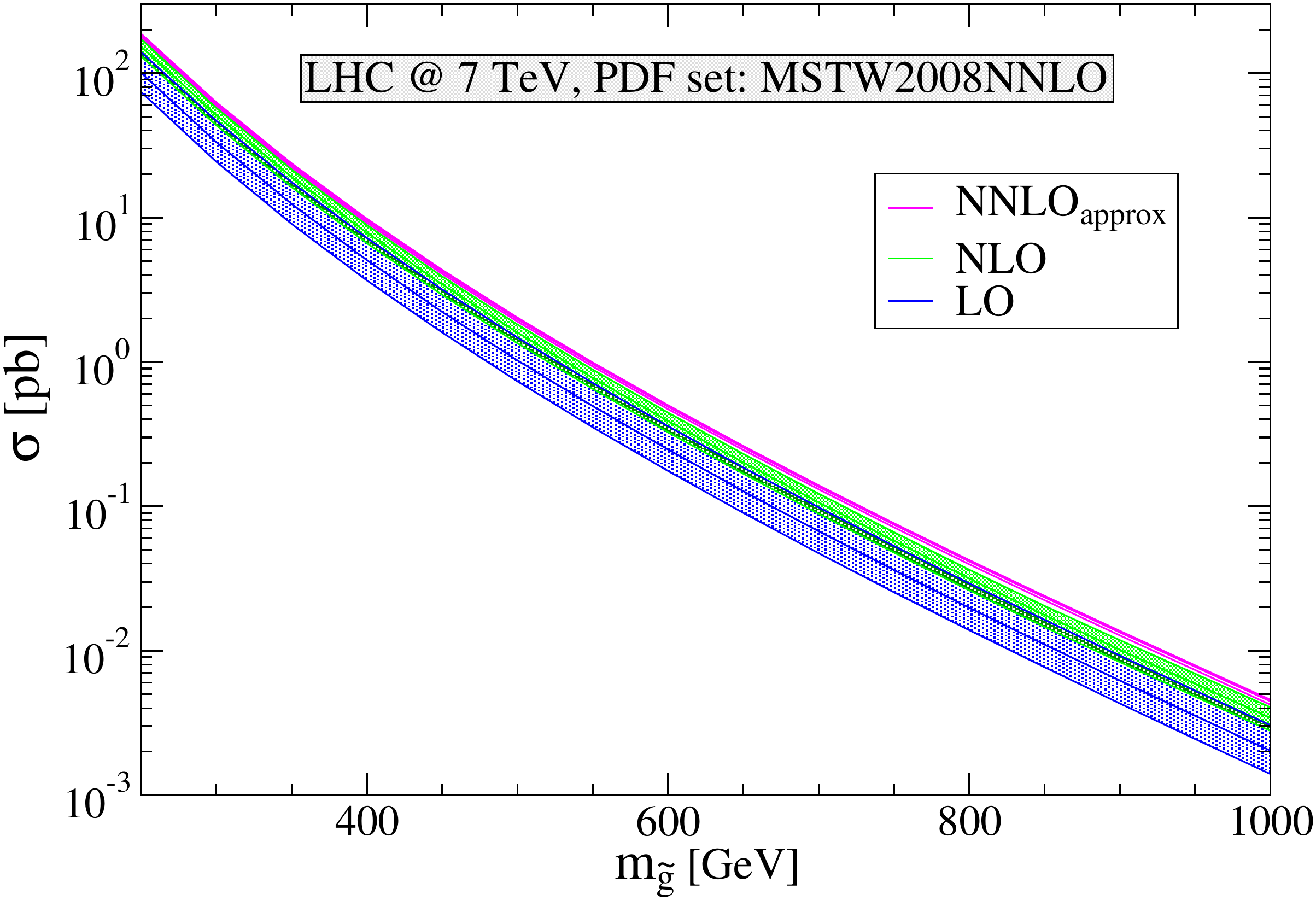}}
\vspace*{15mm}

\scalebox{0.35}{\includegraphics{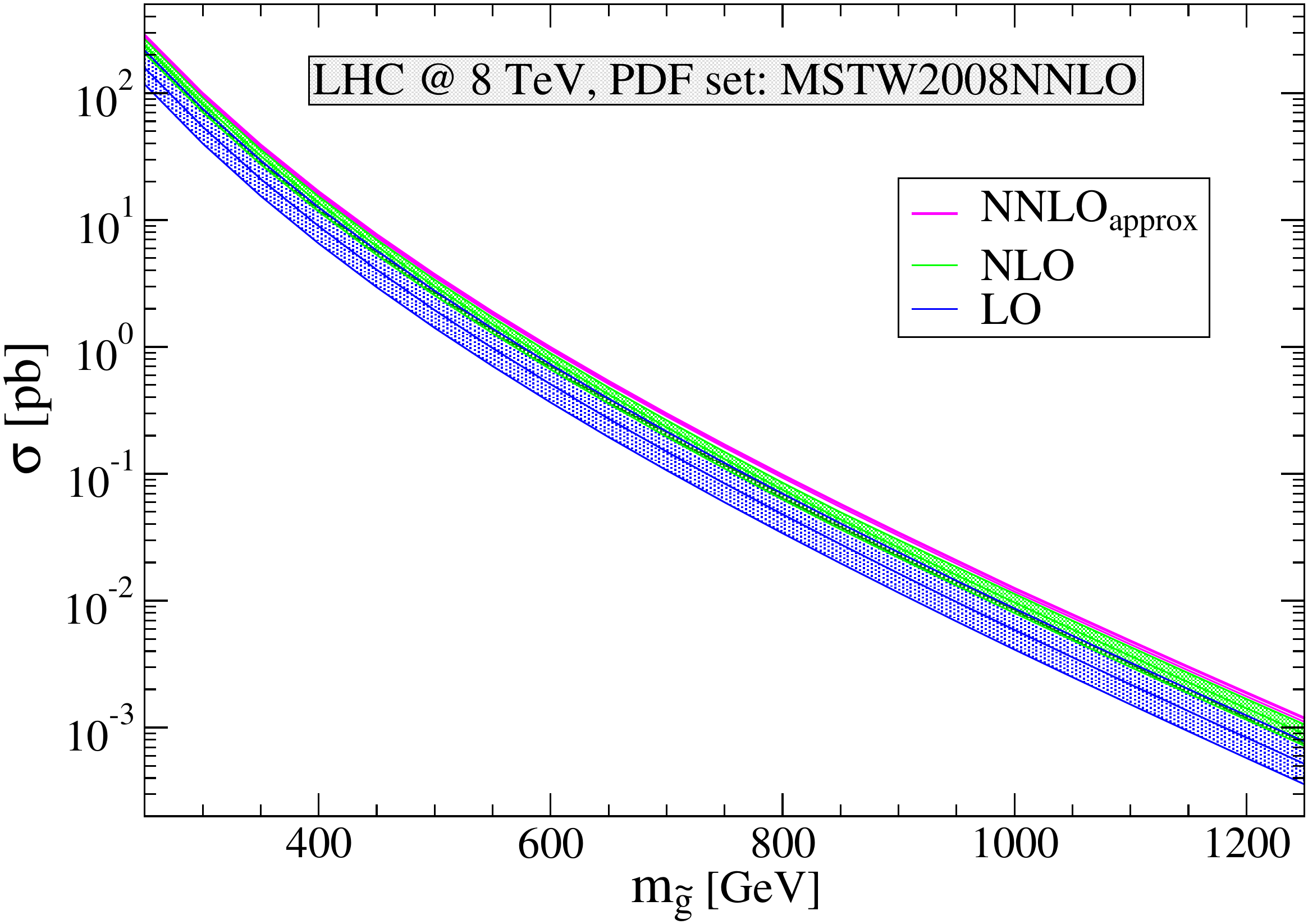}}
\vspace*{15mm}

\scalebox{0.35}{\includegraphics{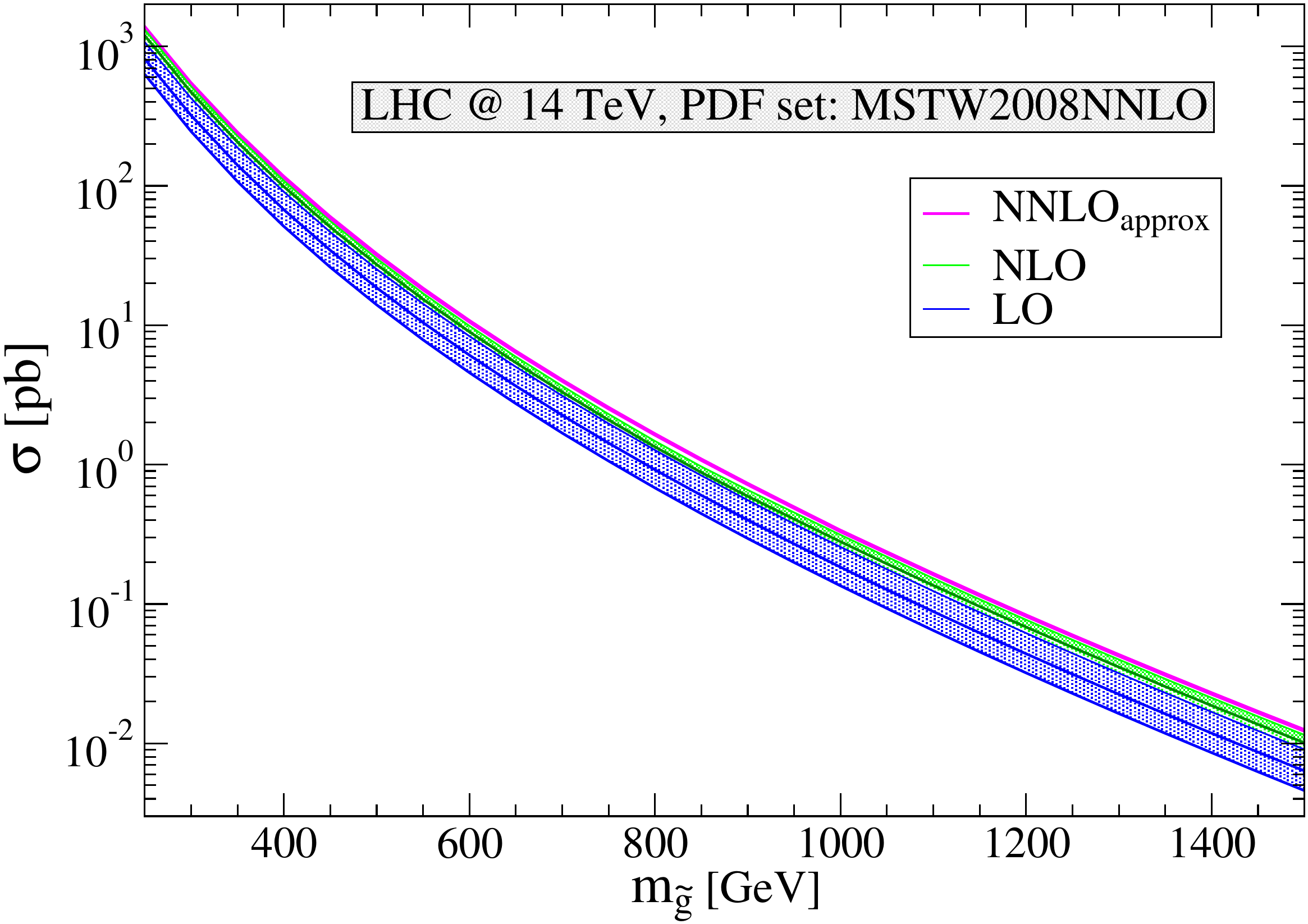}}
\caption{\small{
   Total hadronic cross section and its theoretical uncertainty at 
   the LHC for three different center of mass energies 
   (7\TeV (upper figure), 8 \TeV (central figure), and 14\TeV (lower figure)) 
   at LO (blue bands), NLO (green bands), and NNLO$_{\text{approx}}$ (purple lines) 
   as a function of the gluino mass.
   The masses of the squarks and the stop are set to $\mq = 4/5\mg$.      
   At NNLO$_{\text{approx}}$, the theoretical uncertainty has shrunk to a small band.
}}
\label{fig:shrinkingscaledep}
\end{figure}

We work in the $\overline{\rm MS}$-scheme, which is implemented in 
\texttt{Prospino} 
with $n_l = 5$ light quarks and an on-shell top quark with mass 
$m_t = 175\GeV$~\cite{Beenakker:1996ed}.
The masses of squarks and stops are set equal to the value $\mq = 4/5\mg$ 
so that the gluino is always the heavier particle.
We use the PDF sets ABM11 NNLO~\cite{Alekhin:2012ig} and MSTW2008 NNLO 
PDFs~\cite{Martin:2009iq} irrespective of the order of perturbation theory.
In Fig.~\ref{fig:shrinkingscaledep}, we present total hadronic cross sections for
gluino pair production at the LHC for the cms energies $7\TeV, 8\TeV$ and $14\TeV$
at LO, NLO and NNLO. 
The width of the bands indicates the theoretical uncertainty due to a variation
of the scale $\mu$ in the range $\half \mg \le \mu \le 2\mg$.
The increase in the predicted rates due to the approximate NNLO corrections 
of the order of $\ord(15-20) \%$ at nominal scales is clearly visible 
and cross section numbers for selected gluino masses are given 
in Tabs.~\ref{tab:xsecvalueslhc07} -- \ref{tab:xsecvalueslhc14}.
Over the plotted range of $\mg$, the cross sections in
Fig.~\ref{fig:shrinkingscaledep} 
are decreasing over more than four orders of magnitude. 

\begin{figure}
 \centering
\scalebox{0.5}{\includegraphics{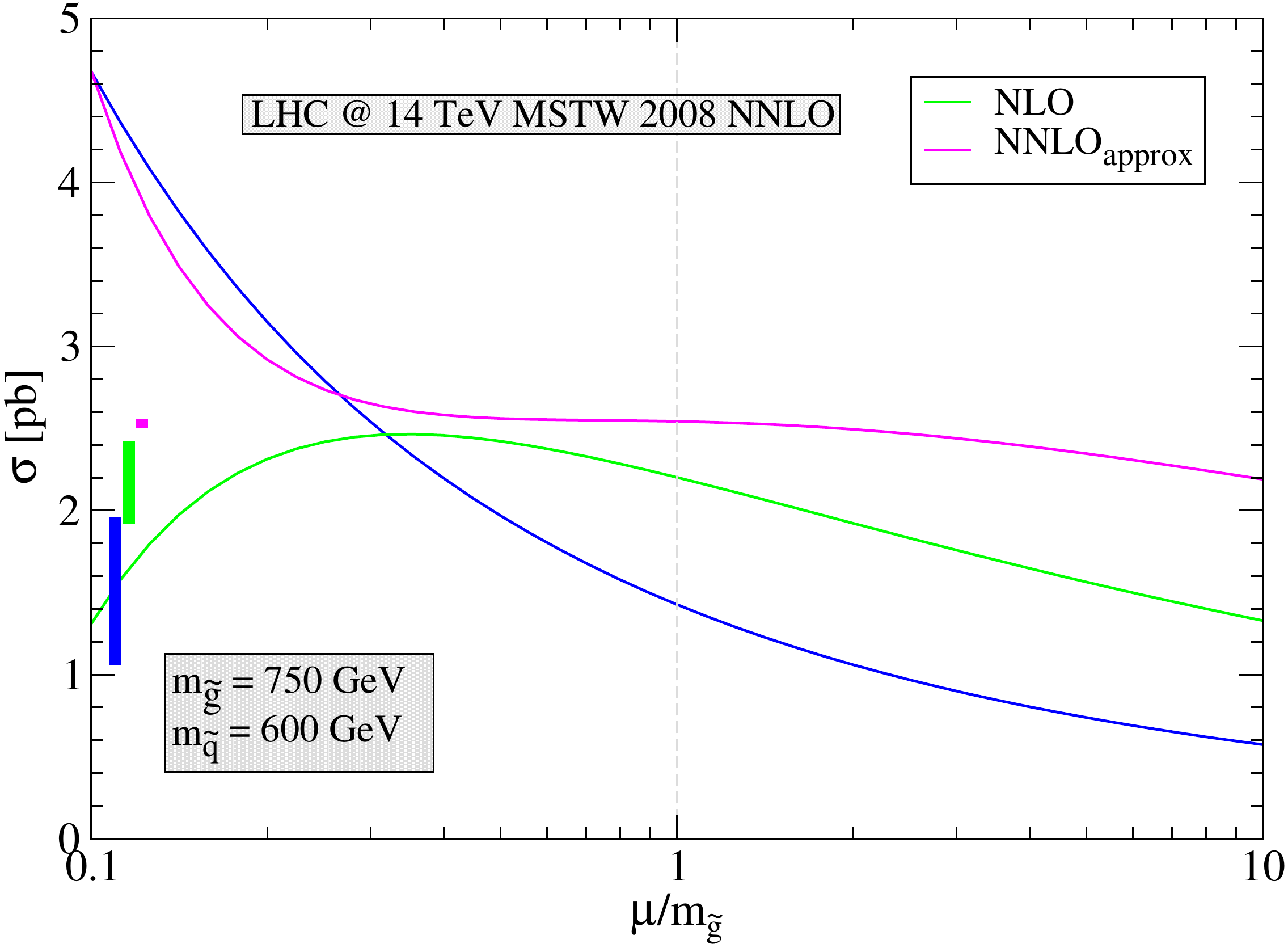}}
\caption{
\small{ 
  Scale dependence of the total hadronic cross section at the LHC with $14\TeV$
  for $\mg = 750\,\GeV$ and $\mq = 600\,\GeV$.
  The vertical bars indicate the total scale variation
  in the range $[\mg/2,2\*\mg]$, the vertical dashed gray line in the middle
  of the figure indicates the cross section at the nominal scale $\mu = \mg$.}
}
\label{fig:mstwscaledep}
\end{figure}

\begin{figure}
 \centering
\scalebox{0.5}{\includegraphics{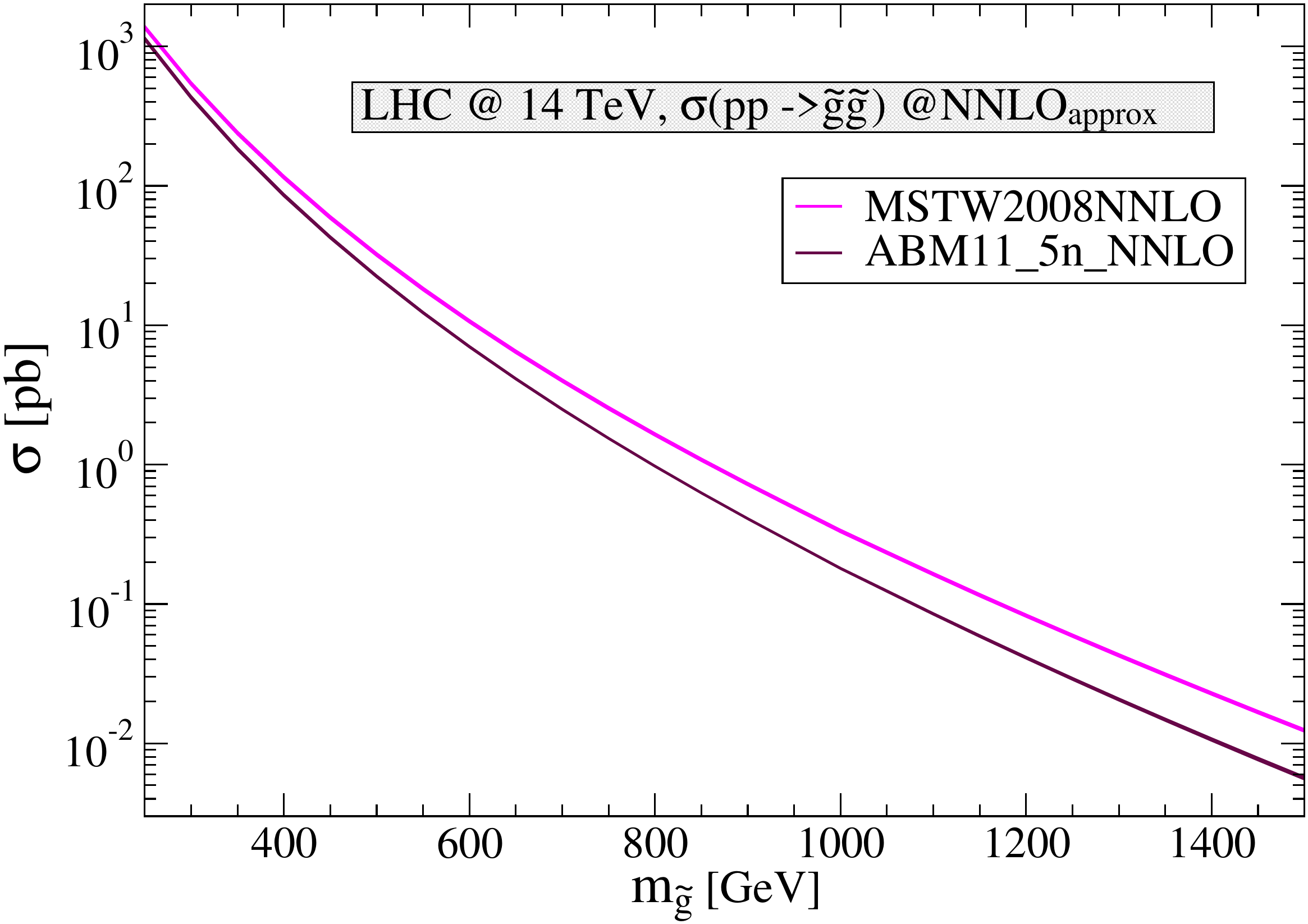}}
\caption{Comparison of the MSTW2008 NNLO~\cite{Martin:2009iq} 
  and ABM11 NNLO~\cite{Alekhin:2012ig} PDF sets.}
\label{fig:abm11nnlovsmstw}
\end{figure}

\begin{figure}
 \centering
\scalebox{0.5}{\includegraphics{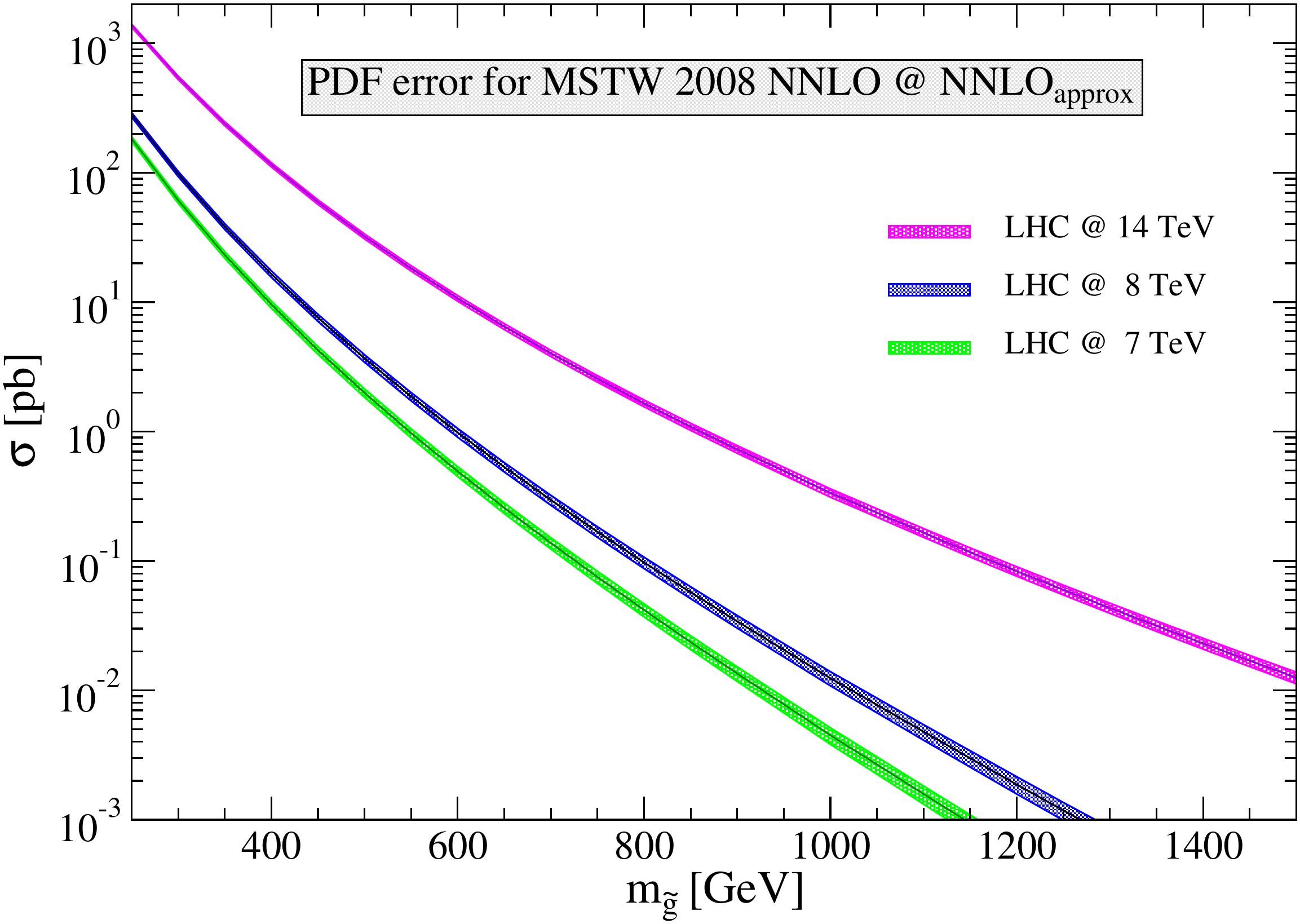}}\\
\vspace{20mm}
\scalebox{0.5}{\includegraphics{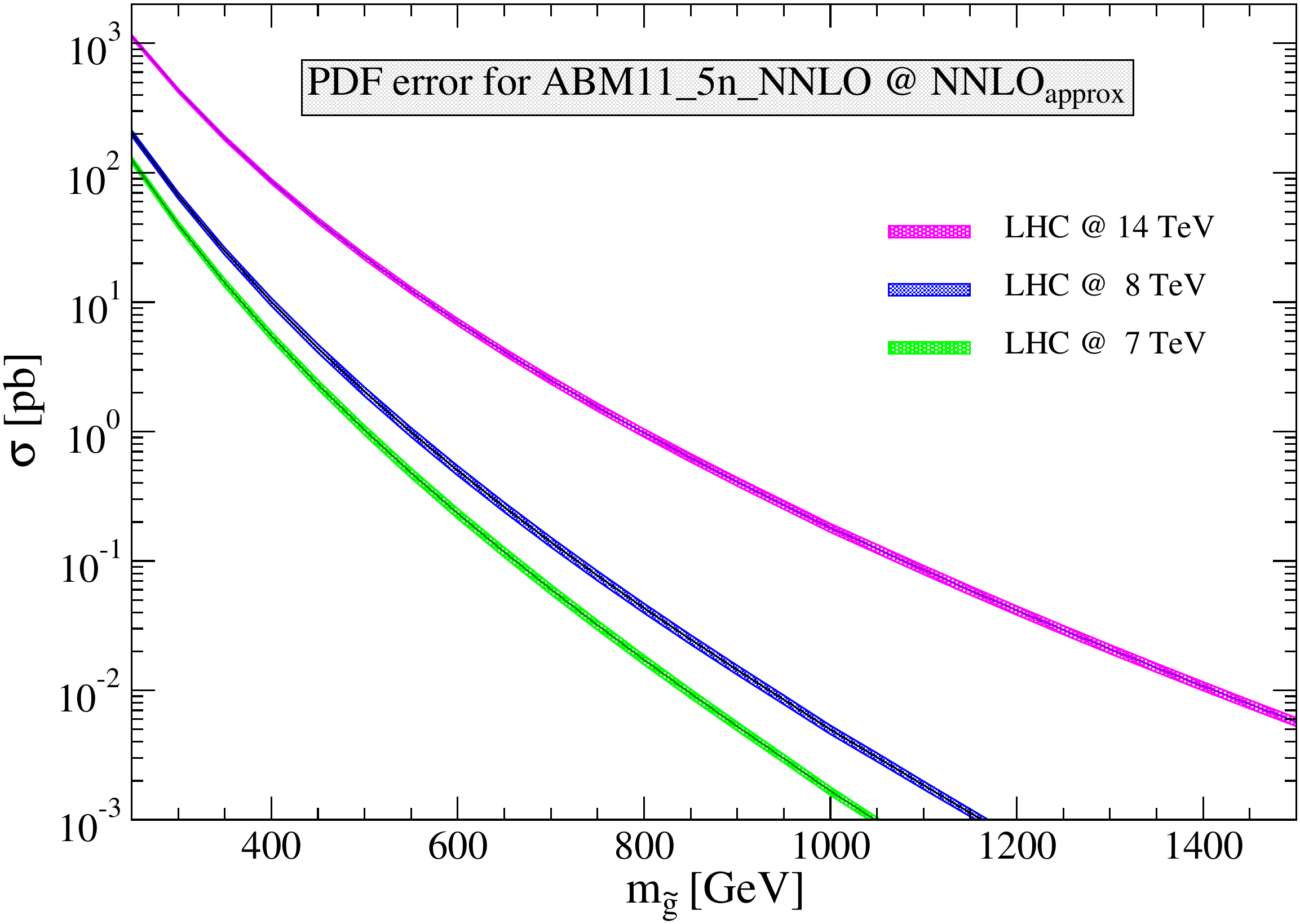}}
\caption{PDF errors of the MSTW2008 NNLO~\cite{Martin:2009iq} 
  and ABM11 NNLO~\cite{Alekhin:2012ig} PDF sets.}
\label{fig:pdferrors}
\end{figure}

In Fig~\ref{fig:mstwscaledep}, we show as an example the scale dependence of the 
hadronic cross section for $\mg = 750\GeV$, $\mq = m_{\tilde{t}_1} = m_{\tilde{t}_2} = 600\GeV$
for the LO, NLO, and approximated NNLO cross section.
The cross section with its uncertainty at the LHC with $14\TeV$ cms 
is $1.43^{+0.53}_{-0.37}\pb$, 
$2.16^{+0.25}_{-0.29}\pb$, and $2.56^{+0.04}_{-0.07}\pb$ at LO, NLO, and NNLO, respectively, where we only quote the errors due to scale variation here.
One observes a strong decrease of the scale uncertainty.
The $K$ factors are $K_{\rm NLO} = \sigma_{\rm NLO}/\sigma_{\rm LO} = 1.46$
and $K_{\rm NNLO} = \sigma_{\rm NNLO}/\sigma_{\rm NLO} = 1.13$ 
and the point of minimal sensitivity, where the cross section adopts similar values for all orders 
is at about $\mu = 0.35\mg$.

In judging these results and the numbers in Tabs.~\ref{tab:xsecvalueslhc07} -- \ref{tab:xsecvalueslhc14}
it should be kept in mind, though, that on top of the scale dependence at NNLO 
there is a residual uncertainty
due to using approximate corrections at NNLO, only.
Depending on the kinematics, 
i.e., the ratio of gluino mass $\mg$ to the hadronic cms energy 
which defines the range for the parton luminosity, this residual uncertainty 
amounts to a few percent ${\cal O}(2-4\%)$, see e.g.,~\cite{Langenfeld:2009wd,Moch:2012mk} 
for estimates obtained in the case of top-quark hadro-production. 
We also mention without discussion that there are
additional uncertainties, e.g., due to the assumption that the
squark spectrum is mass degenerate, which seems unlikely
for a realistic model of nature.

\begin{table}[tbh]
\centering
\begin{tabular}{r|rrr|rrr|rrr}
\toprule
$\mg$ &\multicolumn{3}{c|}{$\sigma(\text{LO}) [\pb]$}&
       \multicolumn{3}{c|}{$\sigma(\text{NLO}) [\pb]$}&
       \multicolumn{3}{c}{$\sigma(\text{NNLO}) [\pb]$}\\
$[\GeV]$& $x = \half$& $x = 1$& $x = 2$& $x = \half$& $x = 1$& $x = 2$& 
          $x = \half$& $x = 1$& $x = 2$\\
\hline
\multicolumn{10}{c}{MSTW 2008 NNLO}\\
\hline
 $   300 $ & $    46.839 $ & $    33.335 $ & $    24.383 $ & $    57.703 $ & $    51.178 $ & $    43.701 $ & $    62.941 $ & $    61.264 $ & $    58.908 $ \\[0mm]
 $   400 $ & $     7.230 $ & $     5.080 $ & $     3.673 $ & $     8.986 $ & $     7.919 $ & $     6.707 $ & $     9.814 $ & $     9.591 $ & $     9.207 $ \\[0mm]
 $   500 $ & $     1.475 $ & $     1.026 $ & $     0.735 $ & $     1.855 $ & $     1.624 $ & $     1.365 $ & $     2.029 $ & $     1.990 $ & $     1.906 $ \\[0mm]
 $   600 $ & $     0.359 $ & $     0.247 $ & $     0.176 $ & $     0.457 $ & $     0.398 $ & $     0.332 $ & $     0.501 $ & $     0.493 $ & $     0.471 $ \\[0mm]
 $   700 $ & $     0.098 $ & $     0.067 $ & $     0.047 $ & $     0.127 $ & $     0.110 $ & $     0.091 $ & $     0.140 $ & $     0.138 $ & $     0.131 $ \\[0mm]
 $   800 $ & $     0.029 $ & $     0.020 $ & $     0.014 $ & $     0.038 $ & $     0.033 $ & $     0.027 $ & $     0.042 $ & $     0.042 $ & $     0.040 $ \\[0mm]
\hline
\multicolumn{10}{c}{ABM11 NNLO}\\
\hline
 $   300 $ & $    29.433 $ & $    21.365 $ & $    15.930 $ & $    36.863 $ & $    32.778 $ & $    28.213 $ & $    40.176 $ & $    39.540 $ & $    38.548 $ \\[0mm]
 $   400 $ & $     4.012 $ & $     2.916 $ & $     2.176 $ & $     5.053 $ & $     4.493 $ & $     3.869 $ & $     5.507 $ & $     5.485 $ & $     5.388 $ \\[0mm]
 $   500 $ & $     0.739 $ & $     0.539 $ & $     0.403 $ & $     0.933 $ & $     0.831 $ & $     0.717 $ & $     1.017 $ & $     1.025 $ & $     1.015 $ \\[0mm]
 $   600 $ & $     0.165 $ & $     0.121 $ & $     0.090 $ & $     0.209 $ & $     0.187 $ & $     0.161 $ & $     0.228 $ & $     0.233 $ & $     0.232 $ \\[0mm]
 $   700 $ & $     0.042 $ & $     0.031 $ & $     0.023 $ & $     0.054 $ & $     0.048 $ & $     0.041 $ & $     0.058 $ & $     0.060 $ & $     0.060 $ \\[0mm]
 $   800 $ & $     0.012 $ & $     0.009 $ & $     0.007 $ & $     0.015 $ & $     0.014 $ & $     0.012 $ & $     0.017 $ & $     0.017 $ & $     0.017 $ \\[0mm]
\bottomrule
 \end{tabular}
    \caption{\small
      \label{tab:xsecvalueslhc07}
      Numerical values for the gluino pair-production cross section
      at LHC with $\sqrt{s}=7\TeV$ and the PDF sets 
      MSTW 2008 NNLO~\cite{Martin:2009iq},
      ABM11 NNLO~\cite{Alekhin:2012ig}.
      The QCD predictions are given at LO, NLO, and NNLO accuracy and
      for different gluino masses and scales $x = \mu/\mg$.}
\end{table}

\begin{table}[tbh]
\centering
\begin{tabular}{r|rrr|rrr|rrr}
\toprule
$\mg$ &\multicolumn{3}{c|}{$\sigma(\text{LO}) [\pb]$}&
       \multicolumn{3}{c|}{$\sigma(\text{NLO}) [\pb]$}&
       \multicolumn{3}{c}{$\sigma(\text{NNLO}) [\pb]$}\\
$[\GeV]$& $x = \half$& $x = 1$& $x = 2$& $x = \half$& $x = 1$& $x = 2$& 
          $x = \half$& $x = 1$& $x = 2$\\
\hline
\multicolumn{10}{c}{MSTW 2008 NNLO}\\
\hline
 $   300 $ & $    75.431 $ & $    54.300 $ & $    40.118 $ & $    92.498 $ & $    82.484 $ & $    70.931 $ & $   100.577 $ & $    97.919 $ & $    94.424 $ \\[0mm]
 $   400 $ & $    12.576 $ & $     8.941 $ & $     6.532 $ & $    15.511 $ & $    13.760 $ & $    11.745 $ & $    16.887 $ & $    16.503 $ & $    15.894 $ \\[0mm]
 $   500 $ & $     2.768 $ & $     1.949 $ & $     1.411 $ & $     3.445 $ & $     3.039 $ & $     2.577 $ & $     3.755 $ & $     3.682 $ & $     3.541 $ \\[0mm]
 $   600 $ & $     0.727 $ & $     0.508 $ & $     0.365 $ & $     0.915 $ & $     0.802 $ & $     0.676 $ & $     0.999 $ & $     0.982 $ & $     0.943 $ \\[0mm]
 $   700 $ & $     0.216 $ & $     0.149 $ & $     0.107 $ & $     0.275 $ & $     0.239 $ & $     0.200 $ & $     0.300 $ & $     0.296 $ & $     0.283 $ \\[0mm]
 $   800 $ & $     0.070 $ & $     0.048 $ & $     0.034 $ & $     0.090 $ & $     0.078 $ & $     0.065 $ & $     0.099 $ & $     0.097 $ & $     0.093 $ \\[0mm]
 $   900 $ & $     0.024 $ & $     0.016 $ & $     0.012 $ & $     0.031 $ & $     0.027 $ & $     0.022 $ & $     0.035 $ & $     0.034 $ & $     0.033 $ \\[0mm]
\hline
\multicolumn{10}{c}{ABM11 NNLO}\\
\hline
 $   300 $ & $    50.103 $ & $    36.595 $ & $    27.433 $ & $    62.375 $ & $    55.683 $ & $    48.137 $ & $    67.796 $ & $    66.555 $ & $    64.864 $ \\[0mm]
 $   400 $ & $     7.444 $ & $     5.435 $ & $     4.071 $ & $     9.320 $ & $     8.313 $ & $     7.182 $ & $    10.132 $ & $    10.047 $ & $     9.853 $ \\[0mm]
 $   500 $ & $     1.486 $ & $     1.087 $ & $     0.815 $ & $     1.868 $ & $     1.667 $ & $     1.441 $ & $     2.030 $ & $     2.034 $ & $     2.008 $ \\[0mm]
 $   600 $ & $     0.360 $ & $     0.264 $ & $     0.198 $ & $     0.453 $ & $     0.405 $ & $     0.351 $ & $     0.493 $ & $     0.499 $ & $     0.495 $ \\[0mm]
 $   700 $ & $     0.100 $ & $     0.073 $ & $     0.055 $ & $     0.126 $ & $     0.113 $ & $     0.098 $ & $     0.137 $ & $     0.140 $ & $     0.140 $ \\[0mm]
 $   800 $ & $     0.030 $ & $     0.022 $ & $     0.017 $ & $     0.038 $ & $     0.034 $ & $     0.030 $ & $     0.042 $ & $     0.043 $ & $     0.043 $ \\[0mm]
 $   900 $ & $     0.010 $ & $     0.007 $ & $     0.006 $ & $     0.013 $ & $     0.011 $ & $     0.010 $ & $     0.014 $ & $     0.014 $ & $     0.014 $ \\[0mm]
\bottomrule
 \end{tabular}
    \caption{\small
      \label{tab:xsecvalueslhc08}
      Same as Tab.~\ref{tab:xsecvalueslhc07} for the LHC with $\sqrt{s}=8\TeV$.}
\end{table}

\begin{table}[tbh]
\centering
\begin{tabular}{r|rrr|rrr|rrr}
\toprule
$\mg$ &\multicolumn{3}{c|}{$\sigma(\text{LO}) [\pb]$}&
       \multicolumn{3}{c|}{$\sigma(\text{NLO}) [\pb]$}&
       \multicolumn{3}{c}{$\sigma(\text{NNLO}) [\pb]$}\\
$[\GeV]$& $x = \half$& $x = 1$& $x = 2$& $x = \half$& $x = 1$& $x = 2$& 
          $x = \half$& $x = 1$& $x = 2$\\
\hline
\multicolumn{10}{c}{MSTW 2008 NNLO}\\
\hline
 $   500 $ & $    25.197 $ & $    18.566 $ & $    13.995 $ & $    30.549 $ & $    27.593 $ & $    24.090 $ & $    32.894 $ & $    32.225 $ & $    31.340 $ \\[0mm]
 $   650 $ & $     5.027 $ & $     3.665 $ & $     2.736 $ & $     6.121 $ & $     5.506 $ & $     4.778 $ & $     6.599 $ & $     6.485 $ & $     6.303 $ \\[0mm]
 $   800 $ & $     1.270 $ & $     0.918 $ & $     0.680 $ & $     1.557 $ & $     1.395 $ & $     1.204 $ & $     1.681 $ & $     1.656 $ & $     1.609 $ \\[0mm]
 $   950 $ & $     0.375 $ & $     0.269 $ & $     0.198 $ & $     0.464 $ & $     0.414 $ & $     0.355 $ & $     0.501 $ & $     0.495 $ & $     0.480 $ \\[0mm]
 $  1100 $ & $     0.124 $ & $     0.088 $ & $     0.065 $ & $     0.154 $ & $     0.137 $ & $     0.117 $ & $     0.167 $ & $     0.165 $ & $     0.160 $ \\[0mm]
 $  1250 $ & $     0.044 $ & $     0.031 $ & $     0.023 $ & $     0.056 $ & $     0.049 $ & $     0.042 $ & $     0.060 $ & $     0.060 $ & $     0.058 $ \\[0mm]
 $  1400 $ & $     0.017 $ & $     0.012 $ & $     0.009 $ & $     0.021 $ & $     0.019 $ & $     0.016 $ & $     0.023 $ & $     0.023 $ & $     0.022 $ \\[0mm]

\hline
\multicolumn{10}{c}{ABM11 NNLO}\\
\hline
 $   500 $ & $    17.216 $ & $    12.852 $ & $     9.815 $ & $    21.146 $ & $    19.120 $ & $    16.772 $ & $    22.759 $ & $    22.453 $ & $    22.042 $ \\[0mm]
 $   650 $ & $     3.127 $ & $     2.332 $ & $     1.778 $ & $     3.860 $ & $     3.486 $ & $     3.055 $ & $     4.155 $ & $     4.131 $ & $     4.075 $ \\[0mm]
 $   800 $ & $     0.729 $ & $     0.544 $ & $     0.415 $ & $     0.903 $ & $     0.816 $ & $     0.715 $ & $     0.972 $ & $     0.974 $ & $     0.966 $ \\[0mm]
 $   950 $ & $     0.201 $ & $     0.150 $ & $     0.115 $ & $     0.250 $ & $     0.226 $ & $     0.198 $ & $     0.269 $ & $     0.272 $ & $     0.271 $ \\[0mm]
 $  1100 $ & $     0.063 $ & $     0.047 $ & $     0.036 $ & $     0.078 $ & $     0.070 $ & $     0.062 $ & $     0.084 $ & $     0.085 $ & $     0.085 $ \\[0mm]
 $  1250 $ & $     0.021 $ & $     0.016 $ & $     0.012 $ & $     0.027 $ & $     0.024 $ & $     0.021 $ & $     0.029 $ & $     0.029 $ & $     0.029 $ \\[0mm]
 $  1400 $ & $     0.008 $ & $     0.006 $ & $     0.004 $ & $     0.010 $ & $     0.009 $ & $     0.008 $ & $     0.010 $ & $     0.011 $ & $     0.011 $ \\[0mm]
\bottomrule
 \end{tabular}
    \caption{\small
      \label{tab:xsecvalueslhc14}
      Same as Tab.~\ref{tab:xsecvalueslhc07} for the LHC with $\sqrt{s}=14\TeV$.}
\end{table}

Finally, we compare the total cross section for the two PDF sets
MSTW2008 NNLO~\cite{Martin:2009iq} and ABM11 NNLO~\cite{Alekhin:2012ig} 
in Fig.~\ref{fig:abm11nnlovsmstw}.
These PDF sets obtained in global fits differ significantly in the value of 
the strong coupling constant $\alpha_s$ and the shape of the gluon PDF at
large parton momentum fraction $x$, 
e.g., $\alpha_s(M_Z) = 0.1134 \pm 0.0011$ for ABM11 and $\alpha_s(M_Z) = 0.1171 \pm 0.0014$ for MSTW.
The differences are marginally compatible,  even if PDF errors are
taken into account. These are plotted in Fig.~\ref{fig:pdferrors} for the cms energies $7$, $8$, and $14\TeV$. Setting $\mu=\mg$, and choosing our default values $\mg=750\GeV$ and $\mq=600\GeV$, we obtain 
$1.55\pm 0.11\pb$, $0.077\pm 0.007\pb$, and $0.032\pm 0.003 \pb$ for ABM11, and
$2.56^{+0.14}_{-0.15}\pb$, $0.168^{+0.026}_{-0.016}\pb$, and
$0.075^{+0.008}_{-0.008}\pb$ for MSTW. 
The origin of these PDF differences 
has been discussed for instance in Ref.~\cite{Alekhin:2012ig}.
As a result, the cross sections calculated with the ABM11 set are 
of the order of $\ord(30-60) \%$ smaller over the whole range of gluino
masses, see also Tabs.~\ref{tab:xsecvalueslhc08} and \ref{tab:xsecvalueslhc14}.
As it stands, the differences in these non-perturbative parameter are the 
largest residual uncertainty in $\gluinopair$-cross section predictions 
with direct implications also for exclusion limits on $\mg$ and $\mq$
reported by the LHC experiments.

\section{Conclusion and Summary}
\label{sec:sum}
We have studied the QCD corrections for gluino pair production at hadron colliders at NNLO in QCD.
With the computation of the hard matching coefficients at NLO 
based on recent results for the production of gluino-bound states~\cite{Kauth:2011vg}, 
we were able to derive all logarithmically enhanced terms near threshold at NNLO.
Our results allow for the evaluation of the resummed $\gluinopair$ cross
section to NNLL accuracy or, alternatively, for predictions 
at approximate NNLO accuracy at fixed order in perturbation theory.
We have chosen the latter approach to illustrate the impact of our new results
on the apparent convergence and the scale stability of the hadronic cross
sections at the LHC.
In summary, we were able to promote the predictions for the gluino pair
production cross section in the threshold region to the next level of
accuracy, now putting it on par with squark-antisquark pair production.

In advancing from NLO to approximate NNLO QCD predictions, we have found a 
significant increase in the rates, with $K$-factors of the order of $\ord(15-20) \%$ 
depending, of course, on the chosen squark and gluino masses. 
The residual scale uncertainty on the other hand is generally small, of the order
of a few percent only, showing good perturbative stability of the result.
The largest uncertainty in the current predictions for $\gluinopair$
hadro-production is due to the necessary non-perturbative input, 
i.e., the value of $\alpha_s(M_Z)$ and the shape of the gluon PDF, where 
differences between the PDF sets ABM11 and MSTW amount to the order of 
$\ord(30- 60) \%$. 
The impact of the latter differences on squark and gluino searches at the LHC 
is dramatic and the implications for any exclusion limits on squark and gluino masses has not 
been addressed so far in experimental analysis.

\subsection*{Acknowledgments}
We thank P.~Marquard for discussions. 
U.L. acknowledges partial support 
by the Helmholtz Alliance {\it ``Physics at the Terascale''} (HA-101) and 
by the Deutsche Forschungsgemeinschaft in Graduiertenkolleg GRK 1147.
S.M. and T.P. have been supported in part 
by the Deutsche Forschungsgemeinschaft in Sonderforschungs\-be\-reich/Transregio~9 and 
by the European Commission through contract PITN-GA-2010-264564 ({\it LHCPhenoNet}).


\begin{appendix}

\renewcommand{\theequation}{\ref{sec:npoint}.\arabic{equation}}
\setcounter{equation}{0}
\section{Scalar n-point functions}
\label{sec:npoint}
Here, we give explicit expressions for the one-, two-, and three-point integrals 
defined in Ref.~\cite{Kauth:2011vg}:
\begin{eqnarray}
a_1(r) &=& r(1 + 2\ln(2) - \ln(r))\,,
 \\[2mm]
b_1(r) &=& 2 + 2\ln(2) - r\ln(r)-(1-r)\,\ln|1-r|\,,
 \\
b_2(r) &=& 2 + 2\ln(2) - \ln(r) + \,\rm{Re}[\,\sqrt{1-r}\,
              \ln(-(1-\sqrt{1-r})(1+\sqrt{1-r})^{-1})\,]\,,
 \\
b_3(r) &=& 1 + 2\ln(2) + r(1-r)^{-1}\ln(r)\,,
 \\
b_4(r) &=& 2 + 2\ln(2) + r\ln(r)-(1+r)\ln(1+r)\,,
 \\
b_5(r) &=& 2 + 2\ln(2) - \ln(r)\,,
 \\
b_6(r) &=& 2 + 2\ln(2) + (r^{-1} -1)\,\ln|1-r|\,,
 \\
b'_1(r) &=& -1 - r\,\ln|(1 - r)/r|\,,
 \\
b'_2(r) &=& \frac 12(1 - r)^{-1} + r\ln(r)(1 - r)^{-2}(1 + r)^{-1}\,,
 \\[2mm]
c_1(r) &=& \frac 14\,\rm{Re}[\, -\Li((5 - r)(3 + 2\sqrt{1-r} - r)^{-1}) + 
                  \Li((1 - r)(3 + 2\sqrt{1-r} - r)^{-1}) \non
 \\
       &&\qquad\quad   - \rm{Li}_2((5 - r)(3 - 2\sqrt{1-r} - r)^{-1}) + 
                  \Li((1 - r)(3 - 2\sqrt{1-r} - r)^{-1})
 \non\\
       &&\qquad\quad   + \Li(4\,(5 + (-2 + r)r)^{-1}) -
                  \Li(4r\,(5 + (-2 + r)r)^{-1})
 \non\\
       &&\qquad\quad   + \ln(1 + (1 - r)^2/4)(\ln(-5 + r) - \ln(-1 + r))
 \non\\
       &&\qquad\quad   - \ln(1 - 4r (5 + (-2 + r)r)^{-1})\ln(r)\,]\,,\\
c_2(r) &=& \frac 12\,\rm{Re}[\,-\Li((3 - r)(1 + r)^{-1}) + 
                  \Li((1 - r)(1 + r)^{-1})
 \\ 
       &&\qquad\quad    + \Li((-3 + r)(1 + r)^{-1}) -
                  \Li((-1 + r)(1 + r)^{-1})\,]\,,
 \non\\
c_3(r) &=& -\frac 18\,\pi^2 + \frac 12\,(\Li((-1 + r)(1 + r)^{-1}) - \Li((1 - r)(1 + r)^{-1}))\,,
 \\
c_4(r) &=& -(1+r)^{-1}( 1 - \ln(2) + \ln(1+r) - (1+r)^{-1}r\ln(r) )\,,
 \\
c_5(r) &=& \frac 12\,\rm{Re}[\,\Li(-1/r) - \Li(1/r)\,]\,. 
\end{eqnarray}
Note that in the limit $r\to 1$, all $c_i(r)$ 
apart from $c_4(r)$ simplify to $-\pi^2/8$.

\renewcommand{\theequation}{\ref{sec:gi}.\arabic{equation}}
\setcounter{equation}{0}
\section{Explicit expressions for the resummation formula}
\label{sec:gi}
Here we give the process-dependent matching constants 
of the general resummation formula~(\ref{eq:sigmaNres}) 
in the $\overline{\rm MS}(n_l=5)$-scheme at NNLL accuracy.
To that end, we find it convenient to introduce the parameter 
$\widetilde N = N\exp(\gamma_E)$ 
and rearrange the terms in Eq.~(\ref{eq:GNexp}) according to
\begin{eqnarray}
  \label{eq:GNexptilde}
  G_{ij,\, {\bf I}}(N) =
  \lnNt \cdot g^1_{ij}(\widetilde\lambda)  +  g^2_{ij,\, {\bf I}}(\widetilde\lambda)  +
  a_s\, g^3_{ij,\, {\bf I}}(\widetilde\lambda)  + \dots\, ,
\end{eqnarray}
with $\widetilde\lambda = a_s\,\beta_0\, \ln \widetilde N$, so that there are no
terms proportional to Euler's constant $\gamma_E$ 
contained in the final results of the coefficients $g^k_{ij,\, {\bf I}}(\widetilde\lambda)$.
With these conventions, we find for the hard constant 
\begin{eqnarray}
\label{eq:g0ggMS}
g^0_{gg,\,{\bf I}} &=& 
1 + a_s\*\bigg{\lbrace}4\*C_{1,\,{\bf I}}^{gg} - 192 
  + C_{\bf I}\*\left(-8 - 4\*\ln(2)\right) + 24\*\ln^2(2) 
  + 12\*\pi^2\bigg{\rbrace}
 \\
&& + a_s^2\*\bigg{\lbrace}16\*C_{1,\,{\bf I}}^{gg}\*\left(-48 
  + 2\*C_{\bf I}\*\left(-1 + \ln(2)\right) + 72\*\ln(2) 
  - 48\*\ln^2(2) + 3\*\pi^2\right)
 \non\\
&& + C_{\bf I}\*\biggl(\frac{8}{3}\*\left(-823 + 1465\*\ln(2)\right) 
  - 4000\*\ln^2(2) + 2496\*\ln^3(2) + 126\*\pi^2
 \non\\
&& - 344\*\ln(2)\*\pi^2 + \left(1296 + 48\*\ln(2)\right)\*\zeta_3\biggr) + 
    C_{\bf I}^2\*\left(32\*\ln(2) - 64\*\ln^2(2) + 4\*\pi^2\right)
 \non\\
&& + \frac{4}{9}\*n_l\*\biggl(C_{\bf I}\*\left(92 - 116\*\ln(2) 
  + 48\*\ln^2(2) - 3\*\pi^2\right) + 1744 - 2560\*\ln(2)
 \non\\
&& + 1776\*\ln^2(2) - 624\*\ln^3(2) - 102\*\pi^2 
  + 108\*\ln(2)\*\pi^2 - 336\*\zeta_3\biggr)
 \non\\ 
&& + \frac{32}{3}\*\left(-8207 + 12260\*\ln(2)\right) - 101344\*\ln^2(2) 
  + 66784\*\ln^3(2) - 23040\*\ln^4(2)
 \non\\
&& + 3292\*\pi^2 - 7992\*\ln(2)\*\pi^2 + 5664\*\ln^2(2)\*\pi^2 + 
    204\*\pi^4 + \left(35728 - 49392\*\ln(2)\right)\*\zeta_3
 \non\\
&& + L_\mu\*\biggl(192\*C_{1,\,{\bf I}}^{gg}\*\left(-1 + \ln(2)\right) + 
       C_{\bf I}\*\left(-472 + 664\*\ln(2) - 576\*\ln^2(2) + 48\*\pi^2\right)
 \non\\
&& - \frac{8}{3}\*n_l\*\left(-68 + 2\*C_{\bf I}\*\left(-1 + \ln(2)\right) 
  + 92\*\ln(2) - 48\*\ln^2(2) + 3\*\pi^2\right) - 21616
 \non\\
&& + 31888\*\ln(2) - 26304\*\ln^2(2) + 10368\*\ln^3(2) + 1332\*\pi^2 - 
       1776\*\ln(2)\*\pi^2 + 8064\*\zeta_3\biggr)
 \non\\
&& + L_\mu^2\*\biggl(-16\*\left(-1 + \ln(2)\right)\*n_l - 2568 
   + 2568\*\ln(2) - 1152\*\ln^2(2) + 144\*\pi^2\biggr)
\bigg{\rbrace} 
 \nonumber
\, ,
\end{eqnarray}
in the case of gluon fusion. 
For the $\qqbar$~channel, we obtain  
\begin{eqnarray}
  \label{eq:g0qqMS}
g^0_{\qqbar,\,{\bf 8}_a} &=& 
1 + a_s\*\bigg{\lbrace} 4\*C_{1,\,{\bf 8}_a}^{\qqbar} + 
\frac{4}{3}\*\left(-82 - 9\*\ln(2) 
  + 8\*\ln^2(2) + 4\*\pi^2\right) 
  + L_\mu\*\left(14 - \frac{4}{3}\*n_l\right)\bigg{\rbrace}
 \\
&& + a_s^2\*\bigg{\lbrace} -\frac{32}{3}\*C_{1,\,{\bf 8}_a}^{\qqbar}\*
    \left(41 - 57\*\ln(2) + 32\*\ln^2(2) - 2\*\pi^2\right)  
  + \frac{4}{81}\*n_l\*\biggl(9460 - 13372\*\ln(2)
 \non\\
&& + 8400\*\ln^2(2) - 2496\*\ln^3(2) - 489\*\pi^2 + 
    432\*\ln(2)\*\pi^2 - 1344\*\zeta_3\biggr) - 29920\*\ln^2(2)
 \non\\
&& + \frac{8}{27}\*\left(-83453 + 127247\*\ln(2)\right) + 
    \frac{2}{9}\*\biggl(79424\*\ln^3(2) - 20480\*\ln^4(2) + 5167\*\pi^2
 \non\\
&& - 10428\*\ln(2)\*\pi^2 + 5248\*\ln^2(2)\*\pi^2 + 168\*\pi^4\biggr) 
  + \frac{16}{3}\*\left(1793 - 1849\*\ln(2)\right)\*\zeta_3
 \non\\
&& + L_\mu\*\biggl(\frac{256}{3}\*C_{1,\,{\bf 8}_a}^{\qqbar}\*\left(-1 + \ln(2)\right) 
  + \frac{16}{27}\,\*n_l\*\left(409 - 553\*\ln(2) + 288\*\ln^2(2) 
  - 18\*\pi^2\right)
 \non\\
&& + \frac{8}{9}\*\Bigl(-8283 + 11819\*\ln(2) - 8640\*\ln^2(2) + 2304\*\ln^3(2) 
  + 502\*\pi^2 - 408\*\ln(2)\*\pi^2
 \non\\ 
&& + 1792\*\zeta_3\Bigr)\biggr) 
  - L_\mu^2\*\frac{32}{9}\*\biggl(10\*n_l\*\left(-1 + \ln(2)\right) + 245 
  + 64\*\ln^2(2) - 245\*\ln(2) -8\*\pi^2\biggr)
\bigg{\rbrace} 
 \non
\, .
\end{eqnarray}
The one-loop matching constants $C_{1,\,{\bf I}}^{gg\,\overline{\rm MS}}$ and 
$C_{1,\,{\bf 8}_a}^{\qqbar\,\overline{\rm MS}}$
are given in Eqs.~(\ref{eq:C1ggI-MS}) and~(\ref{eq:C1qq8-MS}) and 
the variables $C_{\bf I}$ and $D_{\bf I}$ in Eqs.~(\ref{eq:CI}) and~(\ref{eq:DI}).
Note that the presently unknown two-loop matching coefficients $C_{2,\,{\bf I}}^{ij}$ 
have been set to zero in Eqs.~(\ref{eq:g0ggMS}) and (\ref{eq:g0qqMS}).

The Coulomb corrections depend on the color configuration of the
gluino pair, but are independent of the initial state. The matching constant depends
on the Mellin moment $N$. So does the constant for the non-relativistic
kinetic energy correction, which is not related to the resummation of threshold
logarithms. For brevity, we also include this non-Coulomb spin-dependent interaction 
into $g^{0,\,C}_{ij,\,{\bf I}}$. According to Eq.~(\ref{eq:g0fac}), we find
\begin{eqnarray}
  \label{eq:g0C}
  g^{0,\,C}_{ij,\,{\bf I}}(N) &=& 
  1 
  - a_s\,\*4\*D_{\bf I}\*\pi^2\*\sqrt{\frac{N}{\pi}} 
  + a_s^2\,\*\bigg{\lbrace}
  \,\frac{8}{3}\*D_{\bf I}^2\*\pi^4\*N  
  + D_{\bf I}\*\pi^2\*\sqrt{\frac{N}{\pi}}\*\biggl(
  \lnNt\*\Bigl(- 44 + \frac{8}{3}\*n_l\Bigr) \\
&& 
  - \frac{124}{3} + 88\*\ln(2) 
  + \frac{8}{9}\*n_l\*\Bigl(5 - 6\*\ln(2)\Bigr) 
  - L_\mu\*\Bigl(44 - \frac{8}{3}\*n_l\Bigr) \biggr) \non \\
&&
  + 16\*\pi^2\* D_{\bf I} \* \Bigl(3-2\* D_{\bf I}\* (1 + v_{\rm spin})\Bigr)
  \*\Bigl(1-\ln(2)-\frac 12\*\lnNt\Bigr) \bigg{\rbrace} \non
\, .
\end{eqnarray}

For completeness, we also give the coefficients $g^k_{ij,\, {\bf I}}(\widetilde\lambda)$
in the convention of Eq. (\ref{eq:GNexptilde}). Introducing the abbreviations
$L_{fr} = \ln(\mu_f^2/\mu_r^2)$ and
$L_{\widetilde g r} = \ln(4\mg^2/\mu_r^2)$,
we obtain 
\begin{align}
g^1_{ii} &=
 A_i^{(1)}\*\beta_0^{-1}\*\Bigl(
          2
          - 2\*\ln(1 - 2\*\lat)
          + \ln(1 - 2\*\lat)\*\lat^{-1}
          \Bigr)
\,,\\[2mm]
g^2_{ii,\,{\bf I}}  &=
 A_i^{(1)}\*\beta_0^{-3}\*\beta_1\*\Bigl(
          2\*\lat
          + \ln(1 - 2\*\lat)
          + \frac{1}{2}\*\ln^2(1 - 2\*\lat)
          \Bigr)\\
  &\ \ \ + A_i^{(1)}\*\beta_0^{-1}\*\Bigl(
          2\*\lat\,\*L_{fr} 
          + \ln(1 - 2\*\lat)\,\*L_{\widetilde g r}
          \Bigr)\non\\
  &\ \ \ + A_i^{(2)}\*\beta_0^{-2}\*\Bigl(
          - 2\*\lat
          - \ln(1 - 2\*\lat)
          \Bigr)  
         + \beta_0^{-1}\*\frac{1}{2}\*\ln(1 - 2\*\lat)\*
           \Bigl(D_i^{(1)} + D_{gg,\,{\bf I}}^{(1)}\Bigr)\non         
\,,\\[2mm]
g^3_{ii,\,{\bf I}}  &=
  A_i^{(1)}\*\beta_0^{-4}\*\beta_1^2\*\left(
          - \frac{1}{2}\,
          - \lat
          - \left(1
          - \frac{1}{1 - 2\*\lat}\right)\*\ln(1 - 2\*\lat)
          + \frac{1}{2\*(1 - 2\*\lat)}\*\left(1+\ln^2(1 - 2\*\lat)\right)
          \right)\non\\
  &\ \ \  + A_i^{(1)}\*\beta_0^{-3}\*\beta_2\*\left(
          - \frac{1}{2}\,
          + \lat
          + \ln(1 - 2\*\lat)
          + \frac{1}{2\*(1 - 2\*\lat)}
          \right)\non\\
  &\ \ \  + A_i^{(1)}\*\beta_0^{-2}\*\beta_1\,\*L_{\widetilde g r}\*\left(
          - 1
          + \frac{1}{1 - 2\*\lat}\*\left(1 
          + \ln(1 - 2\*\lat)\right)
          \right)\non\\
  &\ \ \  - A_i^{(1)}\*\frac{\pi^2}{6}\*\left(
          2
          - \frac{2}{1 - 2\*\lat}
          \right)
          - A_i^{(1)}\*\left(
          \lat\,\*L_{fr}^2
          + L_{\widetilde g r}^2\*\left(\frac{1}{2}        
          - \frac{1}{2\*(1 - 2\*\lat)}\right)
          \right)\non\\
  &\ \ \  + A_i^{(2)}\*\beta_0^{-3}\*\beta_1\*\left(
          \frac{3}{2}
          + \lat
          - \frac{1}{1 - 2\*\lat}\*\left( \frac{3}{2}
          + \ln(1 - 2\*\lat)\right)
          \right)\non\\
  &\ \ \  + A_i^{(2)}\*\beta_0^{-1}\*\left(
          2\*\lat\,\*L_{fr} 
          + L_{\widetilde g r}\*\left(1
          - \frac{1}{1 - 2\*\lat}\right)
          \right)
          + A_i^{(3)}\*\beta_0^{-2}\*\left(
          - \frac{1}{2}
          - \lat
          + \frac{1}{2\*(1 - 2\*\lat)}
          \right)\non\\
  &\ \ \  + (D_i^{(1)} + D_{gg,\,{\bf I}}^{(1)})\*\beta_0^{-2}\*\beta_1\*\left(
          - \frac{1}{2}\,
          + \frac{1}{2\*(1 - 2\*\lat)}\*\left(1
          + \ln(1 - 2\*\lat)\right)
          \right)\non\\
  &\ \ \ - (D_i^{(1)} + D_{gg,\,{\bf I}}^{(1)})\,\*L_{\widetilde g r}\*\left(
          \frac{1}{2}
          - \frac{1}{2\*(1 - 2\*\lat)}
          \right)
          + (D_i^{(2)} + D_{gg,\,{\bf I}}^{(2)})\*\beta_0^{-1}\*\left(
           \frac{1}{2}\,
          - \frac{1}{2\*(1 - 2\*\lat)}
          \right)
\,.
\end{align}

\end{appendix}

\begin{footnotesize}


\end{footnotesize}

\end{document}